\documentclass[12pt]{elsarticle}

\usepackage{graphicx}

\usepackage{amssymb}

\journal{Journal of Computational Physics}

\newcommand{\be}{\begin{equation}}
\newcommand{\ee}{\end{equation}}
\newcommand{\bea}{\begin{eqnarray}}
\newcommand{\eea}{\end{eqnarray}}
\newcommand{\K}{{\tt KADATH} }
\newcommand{\cinf}{${\mathcal C }^\infty$}
\newcommand{\fraction}{\displaystyle\frac}
\newcommand{\R}{\mathbb R }
\begin{document}

\begin{frontmatter}



\title{KADATH: a spectral solver for theoretical physics}


\author{Philippe Grandcl\'ement}
\ead{philippe.grandclement@obspm.fr}
\address{Laboratoire de l'Univers et Th\'eories LUTH, Observatoire de Paris-Meudon, 5 Place J. Janssen, 91195 Meudon Cedex, France.}

\begin{abstract}
\K is a library that implements spectral methods in a very modular manner. It is designed to solve a wide class of problems that arise in the context of theoretical physics. Several types of coordinates are implemented and additional geometries can be easily encoded. Partial differential equations of various types are discretized by means of spectral methods. The resulting system is solved using a Newton-Raphson iteration. Doing so, \K is able to deal with strongly non-linear situations. The algorithms are validated by applying the library to four different problems of contemporary physics, in the fields of gauge field theory and general relativity.
\end{abstract}


\end{frontmatter}

\section{Introduction}

Spectral methods are a class of numerical methods that aim at solving partial differential equations. For detailed presentations of those techniques, the reader should refer to the numerous books available like \cite{GotllO77, CanutHQZ88, CanutHQZ06, CanutHQZ07, Fornb95, Boyd01}. The basic idea is to describe any field by an appropriate linear combination of known functions called the basis functions. Classic examples of basis are the trigonometrical functions or orthogonal polynomials (Chebyshev, Legendre, etc.). The description of functions by their spectral expansion is by essence non-local, which is to be contrasted with finite difference schemes. Spectral methods are particularly appealing because of the very fast convergence of the spectral expansion to the real function it describes. More precisely for \cinf functions the error decreases faster than any power-law of $N$, where $N$ is the order of the expansion. A multi-domain decomposition, where the physical space is decomposed into several computational regions, is usually used to ensure such smoothness of the functions.

Originally, spectral methods were used in the context of numerical hydrodynamics and led to the successful computations of turbulence regimes and various two or three-dimensional flows. Numerous applications can be found in \cite{CanutHQZ88, CanutHQZ07}. The application of spectral methods to the field of general relativity is somewhat more recent and starts with the pioneer work of the Meudon group, in the late eighties. Since then, such methods have been successfully applied by several groups to systems like binary black holes or magnetized neutron stars. A review of spectral methods in the context of general relativity can be found in \cite{GrandN09}.

{\tt LORENE} is the library written by the Meudon group that implements spectral methods for general relativity. It is written in C++, is used by several groups worldwide and produced many results (see \cite{lorene} for references). Nevertheless, with time, it appeared that {\tt LORENE} did encounter some difficulties and had some serious limitations. The first one comes from the fact that the library is designed to work with spherical coordinates only and from the difficulty to implement new geometries like the bispherical one for instance. Second, {\tt LORENE} solves systems by an iterative loop that requires a splitting of the equations in terms of operators (like the Laplace one) and sources. This splitting is obviously not unique. If it can be natural in some cases, this is not true when strong non-linearities occur, like for gauge field problems where the use of {\tt LORENE} proved problematic (the example shown in Sec. \ref{ss:vortons} was never successfully computed by {\tt LORENE}). Finally, and even if learning {\tt LORENE} is manageable, it can still be somewhat difficult to write a complete code. A more user-friendly coding style would be a good thing.

For all those reasons, it has been decided to think about a new library that would use the many successes of {\tt LORENE} and try to improve on its weaknesses. The \K library is the result of this process and this work is the first paper devoted to its description. The design of \K was also inspired by some aspects of other spectral solvers, like the ones described in \cite{PfeifKST03, AnsorKM03, Ansor05}. Three different types of coordinates have been coded so far but the setting is such that the inclusion of additional geometries is relatively easy and straightforward. Spectral expansion is done either with respect to Chebyshev or Legendre polynomials and trigonometrical functions. Systems of equations are solved as such, by means of a Newton-Raphson iteration and do require only a minimal rewriting of the equations. They are passed to the solver as character strings. Those strings are interpreted by \K using a syntax close to the LateX one \cite{latex}, so that the interface should be relatively clear to any modern physicist. The library is written in C++. It has been made public and can be downloaded by going to the \K web-site \cite{kadath}.

In its current state of development, the library deals with boundary value problems. Moreover, the geometry of the various boundaries must be known in advance and is fixed during the whole computation. Such kind of setting is useful in computing stationary or periodic solutions and encompasses a wide class of problems, as the four examples discussed in Sec. \ref{s:test} illustrate. Nevertheless, the field of application of \K would be significantly broadened if some of those restrictions could be lifted. One possible extensions would be the possibility to deal with free boundary problems where the geometry is no longer fixed but numerically determined in the course of the computation. This would be required, for instance, to compute neutron stars in rotation, where the shape of the surface of the star is an unknown. Another addition to \K would be the inclusion of tools to study time-evolution problems. This is however a major task that would require much time and work. Those possible extensions, as long as a few others, are discussed in more detail in Sec. \ref{s:conclusion}.

This paper contains two main parts. In the first one, the basic numerical techniques used by the library are exposed. Not all the details are given, for it would be cumbersome. Nevertheless it should give the reader a good feeling of what \K is about. Section \ref{s:geometry} describes the way multi-domain settings are implemented along with the three different types of geometries currently present in \K. In Sec. \ref{s:spectral}, after a short introduction on spectral expansion, the various basis used by \K are detailed, for different geometries and different types of mathematical objects (scalars, tensors, etc.). The way additional regularities are enforced is also discussed in some details. Section \ref{s:equations} is devoted to the discretization of  partial differential equations by $\tau$ and Galerkin methods. The resulting non-linear system is solved by a Newton-Raphson iteration which is described in Sec. \ref{s:system}. The computation of the Jacobian and its inversion are discussed, especially in the context of parallel computing. Section \ref{s:user} gives an outline of what a code that uses \K should look like.

The second part of the paper is concerned with four different test problems (Sec. \ref{s:test}). Those problems have been chosen because they illustrate different aspects of \K, in terms of the type of equations, variables or geometries. If those are not exactly new results, they are far from being trivial and they all relate to very contemporary physics. The existence of vortons in a particular gauge field theory is confirmed in \ref{ss:vortons}, a critical collapse solution is constructed in Sec. \ref{ss:critic}, binary black holes spacetimes are obtained in \ref{ss:hkv} and the Kerr solution for a single rotating black hole is computed in \ref{ss:kerr}. Future developments and applications are discussed in Sec. \ref{s:conclusion}.

\section{Setting the geometry}\label{s:geometry}
\subsection{Multi-domain decomposition}\label{ss:multidomain}
\K implements spectral methods in a multi-domain manner where the physical space is split into several computational domains. The advantages of using a multi-domain setting are numerous. First of all, when dealing with discontinuous fields, the domains can be chosen so that the surfaces of discontinuities coincide with the boundaries. Doing so, in each domains, all the functions are \cinf, thus recovering the spectral convergence that would be lost otherwise (see Sec. \ref{ss:spectral}). The use of several domains also enables to use different resolutions in different parts of space, increasing accuracy in regions where needed. This is called fixed-mesh refinement. There are also some cases where setting a global and regular set of numerical coordinates is troublesome. This is for instance the case of the bispherical coordinates, as can be seen in Sec. \ref{ss:bispheric}. Such difficulties can be overcome by setting different numerical coordinates in different regions of space.

For each domain, there is a mapping between a set of {\em physical coordinates} (the $\left(r, \theta, \varphi\right)$ of spherical coordinates for instance) and a set of {\em numerical coordinates}. The spectral expansion is performed with respect to the numerical ones. The mapping between the two sets of coordinates ensures that the numerical ones lies in the appropriate range to do the expansion. For instance, a coordinate must be in $\left[-1, 1\right]$ if Chebyshev polynomials are used.

The physical space is divided into a finite set of domains. Only touching and not overlapping domains are considered. The domains are described by the class {\tt Domain} and its derived classes. The way the various domains are set with respect to each other is encoded in the abstract class {\tt Space} and its derived classes. This is needed, for instance, to write the appropriate matching conditions across the boundaries of the domains.

\K provides several domain decompositions that are listed below. The library has been designed to be very modular in terms of the geometry, so that it is relatively easy to implement new types of domains and spaces. Among the possible future developments, one could think about square-like domains, three-dimensional cylindrical coordinates or spaces with more than three dimensions.

\subsection{Spherical space}\label{ss:spheric}

The class {\tt Space\_spheric} implements a decomposition of space in terms of spherical shells. This is obviously intended mainly for spherical-like objects but has also been successfully used for more complicated shapes, like the toroidal one (see Sec. {\ref{ss:vortons}). This geometry is very similar to the one described in
\cite{BonazGM98}.

In this setting, the physical coordinates are the standard spherical ones $\left(r, \theta, \varphi\right)$. One notes $\theta$ the zenith angle and $\varphi$ the azimuthal one, so that, in terms of Cartesian coordinates, one gets:
\bea
x &=& r \sin\theta \cos\varphi \\
y &=& r \sin\theta \sin\varphi \\
z &=& r \cos\theta
\eea

For all the type of domains, the numerical angular coordinates  $\left(\theta^\star, \varphi^\star\right)$ are identical to their physical counterparts.  So far, the various fields are supposed to be either symmetric or antisymmetric with respect to the $z=0$ plane. This implies that $\theta$ can be restricted to $\left[0, \pi/2\right]$. No symmetry is assumed with respect to $\varphi$, which lies in $\left[0, 2 \pi\right[$. Concerning the radial coordinate $r$, three different types of domains are considered.

\begin{itemize}
\item The class {\tt Domain\_nucleus} represents a spherical domain that encompasses the origin of the coordinate system and that extends up to a finite radius. The numerical radial coordinate $r^\star$ relates to $r$ by $r = \alpha r^\star$, where $\alpha$ is a constant that gives the radius of the domain. $r^\star$ lies in $\left[0, 1\right]$.
\item The class {\tt Domain\_shell} represents a spherical shell, where the radius $r$ lies between to finite values. The numerical coordinate $r^\star$ relates to $r$ by an affine law $r=\alpha r^\star + \beta$, where $\alpha$ and $\beta$ are constants. $r^\star$ spans the intervale $\left[-1, 1\right]$.
\item The last type of spherical domain is called {\tt Domain\_compact} an extends from some finite radius up to infinity. This is done by demanding that $r^\star$ relates to $r$ by $r=\fraction{1}{\alpha\left(r^\star-1\right)}$, where $\alpha$ is a negative constant. $r^\star$ lies in $\left[-1, 1\right]$. As intended, the domain goes to $r=\infty$, which corresponds to $r^\star = 1$.
\end{itemize}

For axisymmetric problems that do not depend on $\varphi$, a variant of the spherical space has been implemented. It is called {\tt Space\_polar} and the only difference is that the variable $\varphi$ has been removed, making the space two-dimensional, thus reducing the computational cost.

\subsection{Bispherical space}\label{ss:bispheric}

Bispherical coordinates are implemented by the class {\tt Space\_bispheric}. In the standard point of view, bispherical coordinates 
$\left(\chi, \eta, \varphi \right)$ relate to the Cartesian ones by:
\bea\label{e:bispheric}
x &=& \frac{a\sinh\eta}{\cosh\eta - \cos \chi}\\
\nonumber y &=& \frac{a \sin\chi \cos\varphi}{\cosh\eta - \cos \chi}\\
\nonumber z &=& \frac{a\sin\chi \sin\varphi}{\cosh\eta - \cos \chi}
\eea

The coordinate $\eta$ can take all the values in $\R$, whereas $\chi$ goes from $0$ to $\pi$. $\varphi$ is the angle around the $x$-axis and it lies in $\left[0, \pi\right]$, once the symmetry with respect to the plane $z=0$ is taken into account.

Bispherical coordinates are such that the surfaces of constant $\eta$ are non-concentric spheres located on the $x$-axis, hence their name. There are given by 
\be
\label{e:spheres}
\left(x-a\coth\eta\right)^2 + y^2 + z^2 = \fraction{a^2}{\sinh^2\eta}
\ee
 This property makes those coordinates very appropriate to deal with physical systems consisting of two spherical-like objects. So far, \K enables the user to consider the space exterior to two spheres (for an application see Sec. \ref{ss:hkv}). Let us call $r_1$ and $r_2$ the radii of those spheres and $d$ their separation. It is then easy to see that the space exterior to those two spheres is described by the points such that $\eta \in \left[\eta^-, \eta^+\right]$. The values of $\eta^\pm$ and the scale factor $a$ are then uniquely defined and obey the following set of equations:
\bea
\sinh \eta^- &=& \frac{a}{r_1} \label{e:r1} \\
\sinh \eta^+ &=& \frac{a}{r_2} \label{e:r2} \\
d &=& a\left(\coth\eta^+ - \coth\eta^-\right) \label{e:distance}.
\eea
Those equations simply translate in terms of $\eta$ the positions and sizes of the spheres by making use of Eq. (\ref{e:spheres}). Let us note that, doing so, the origin of the Cartesian coordinates (i.e. the point $x=y=z=0$) cannot be placed arbitrarily. 

In bispherical coordinates, spatial infinity is described by $\eta=0$ and $\chi=0$ implying that the surface $r=\infty$ is described by a single line. This makes a simple compactification like the one used in the spherical case (see Sec. \ref{ss:spheric}) impracticable. This can be seen if one forms the ratio of the Cartesian coordinates with respect to $r$. As one approaches spatial infinity such ratio goes like
\bea
\frac{x}{r} \rightarrow \frac{\eta}{\sqrt{\eta^2 + \chi^2}}
\eea
which does not admit a well-defined limit when $\eta \rightarrow 0$ and $\chi\rightarrow 0$. This effect is closely related to the fact that the coordinate transformation (\ref{e:bispheric}) si only ${\mathcal C}^0$ and not \cinf at infinity, as can be seen in \cite{Ansor05}. 

One way to overcome this difficulty is to excise a region of space around the point $\eta=0$ and $\chi=0$ (see \cite{Ansor05} for another possibility). Doing so, the physical space does no longer extend up to infinity but only to a finite distance. The shape of the newly created exterior boundary is defined by the shape of the excision region. For instance, in order to get a spherical outer boundary of radius $R$, one needs to excise the region:
\be\label{e:excision}
\frac{\sinh^2\eta + \sin^2\chi}{\left(\cosh\eta - \cos\chi\right)^2} \leq \frac{R^2}{a^2}.
\ee
The presence of the exterior boundary can also be useful for some problems for which outer boundary conditions must be prescribed. This is especially true for problems with radiation where outgoing waves can be present.

To summarize, \K implements bispherical coordinates in the region defined by $\eta^-\leq \eta \leq \eta^+$ and Eq. (\ref{e:excision}). This corresponds to the region of space exterior to two spheres of radii $r_1$ and $r_2$ and distant from $d$ and inside an outer sphere of radius $R$. The center of the outer sphere coincides with the origin of the coordinates and is not freely specifiable. However, if needed, Eq. (\ref{e:excision}) could be modified to accommodate any smooth outer boundary. Both the inside of the two inner spheres and the exterior of the outer one, can be matched with spherical-like domains as those seen in 
Sec. \ref{ss:spheric} to describe the desired physical space.

In order to perform numerical expansions, one needs to map the bispherical coordinates $\left(\eta, \chi\right)$ onto some numerical coordinates 
$\left(\eta^\star, \chi^\star\right)$. Unfortunately, with the region previously defined, this cannot be done simply using a single domain. The bispherical region must be split into several domains, each of them being mapped onto squares of numerical coordinates. In order to do the splitting the following types of domains are defined.
\begin{itemize}
\item The class {\tt Domain\_bispheric\_rect} represents a rectangular domain in terms of $\left(\eta, \chi\right)$. More precisely, $\eta$ lies in between two finite values
$\eta_{\rm min}$ and $\eta_{\rm max}$. It relates to the numerical coordinate $\eta^\star$ by an affine-law
\be\label{e:eta_affine}
\eta = \frac{\eta_{\rm max} - \eta_{\rm min}}{2} \eta^\star + \frac{\eta_{\rm max} + \eta_{\rm min}}{2}
\ee
so that $\eta^\star \in \left[-1, 1\right]$. $\chi$ goes from a lower boundary $\chi_{\rm min}$ up to $\pi$ and relates to its numerical counterpart via
\be
\chi = \left(\chi_{\rm min}-\pi\right) \chi^\star + \pi
\ee
so that $\chi^\star\in\left[0, 1\right]$. This choice of mapping will be justified by the chosen basis of decomposition in Sec. \ref{sss:bispheric_base}.
\item The class {\tt Domain\_bispheric\_chi\_first} describes domains for which $\eta$ is given as a function of $\chi$. $\chi$ goes from $0$ to a boundary 
$\chi_{\rm max}$ and relates to $\chi^\star$ by 
\be
\chi = \chi_{\rm max} \chi^\star
\ee
$\chi^\star$ lies in $\left[0, 1\right]$. $\eta$ goes from fixed value $\eta_{\rm lim}$ up to a variable bound $f\left(\chi\right)$. The function $f\left(\chi\right)$ is given by the surface of excision (\ref{e:excision}) and is computed numerically. $\eta^\star$ goes from $\left[-1, 1\right]$ and is given by
\be
\eta = \frac{f\left(\chi\right) - \eta_{\rm lim}}{2} \eta^\star + \frac{f\left(\chi\right) + \eta_{\rm lim}}{2}
\ee
\item In the class {\tt Domain\_bispheric\_eta\_first}, the variable $\chi$ is given in terms of $\eta$. $\eta$ is located between $\eta_{\rm min}$ and 
$\eta_{\rm max}$ and relates to $\eta^\star$ simply by Eq. (\ref{e:eta_affine}). $\chi$ goes a variable bound $g\left(\eta\right)$ up to $\pi$. As for the domain {\tt Domain\_bispheric\_chi\_first}, the function $g\left(\eta\right)$ is defined by the excision shape. One then gets 
\be
\chi = \left(g\left(\eta\right) - \pi\right) \chi^\star + \pi
\ee
with $\chi^\star \in\left[0, 1\right]$
\end{itemize}

In the three types of domains $\varphi$ coincides with the associated numerical coordinates and goes from $0$ to $\pi$. 

A set of five such domains is needed to describe the space between the two inner spheres and the exterior one. This is done in the following way. Let us pick two sets of values $\left(\eta_1, \chi_1\right)$ and $\left(\eta_2, \chi_2\right)$ that lie on the excision region. This means that they correspond to circles on the exterior sphere. One will assume that $\eta_1 < 0$ and that $\eta_2 > 0$. The domain decomposition is as follows, the labels of the domains are given in Fig. \ref{f:bispheric}:
\begin{itemize}
\item One {\tt Domain\_bispheric\_chi\_first} for which $0\leq \chi \leq \chi_1$ (domain 1).
\item One {\tt Domain\_bispheric\_rect} for which $\chi_1\leq \chi \leq \pi$ and $\eta^- \leq \eta \leq \eta_1$ (domain 2).
\item One {\tt Domain\_bispheric\_eta\_first} for which $\eta_1\leq \eta \leq \eta_2$ (domain 3).
\item One {\tt Domain\_bispheric\_rect} for which $\chi_2\leq \chi \leq \pi$ and $\eta_2 \leq \eta \leq \eta^+$ (domain 4).
 \item One {\tt Domain\_bispheric\_chi\_first} for which $0\leq \chi \leq \chi_2$ (domain 5).
\end{itemize}
As already stated, spherical domains can be added to this decomposition to account for the interior of the inner spheres or for the exterior of the outer one. This setting is shown in Fig. \ref{f:bispheric} both in the $\left(\eta, \chi\right)$ plane and in terms of Cartesian coordinates. Table \ref{t:domains_bi} shows to what correspond the boundaries of each domain. In particular, one can see that the mappings have be chosen so that $\chi^\star=0$ always corresponds to the $x$-axis. This property will prove useful when regularity conditions will be enforced (see Sec. \ref{sss:axis_regularity}).

\begin{figure}[htb]
\includegraphics[width=0.6\textwidth]{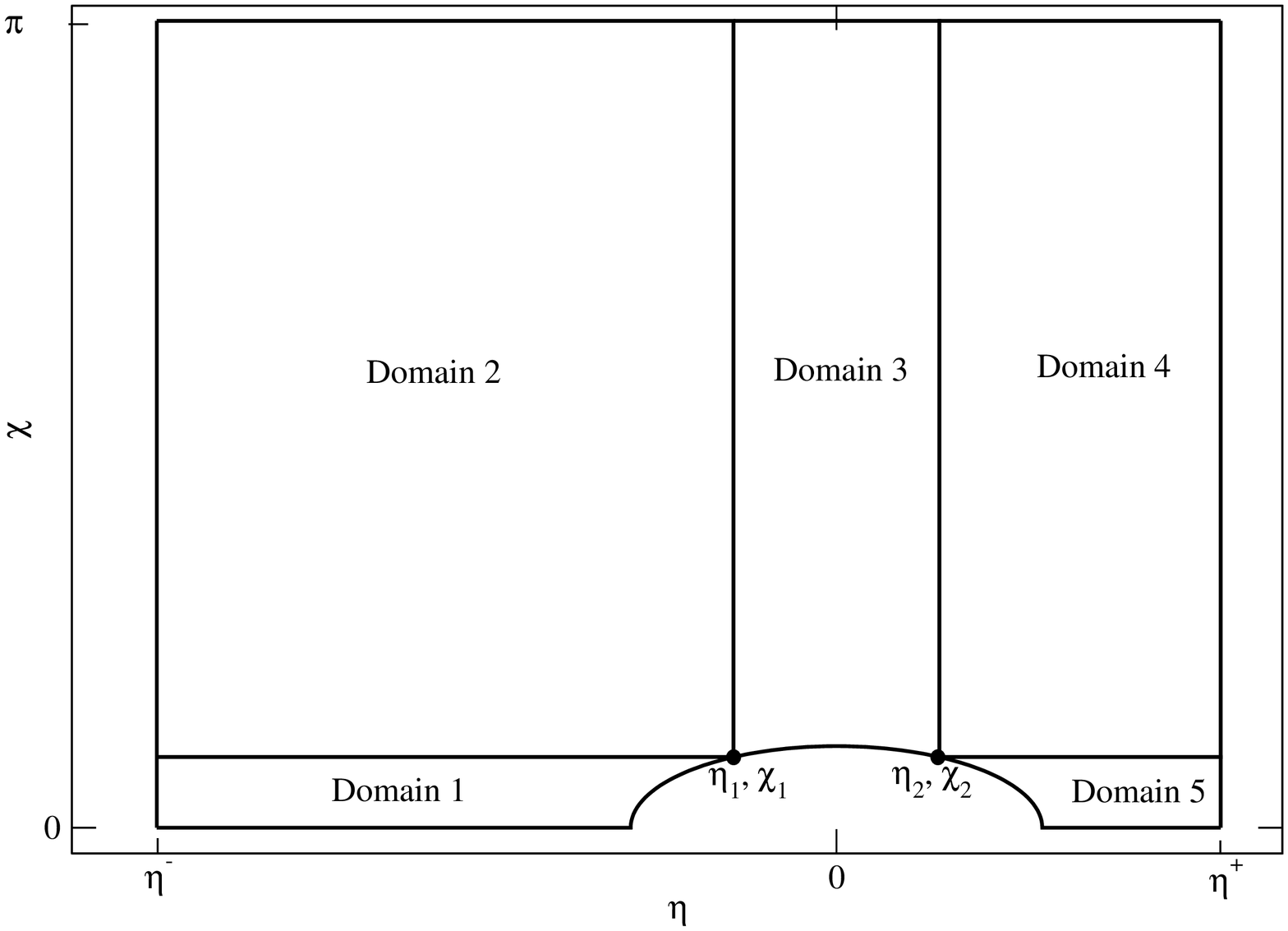} 
\includegraphics[width=0.6\textwidth]{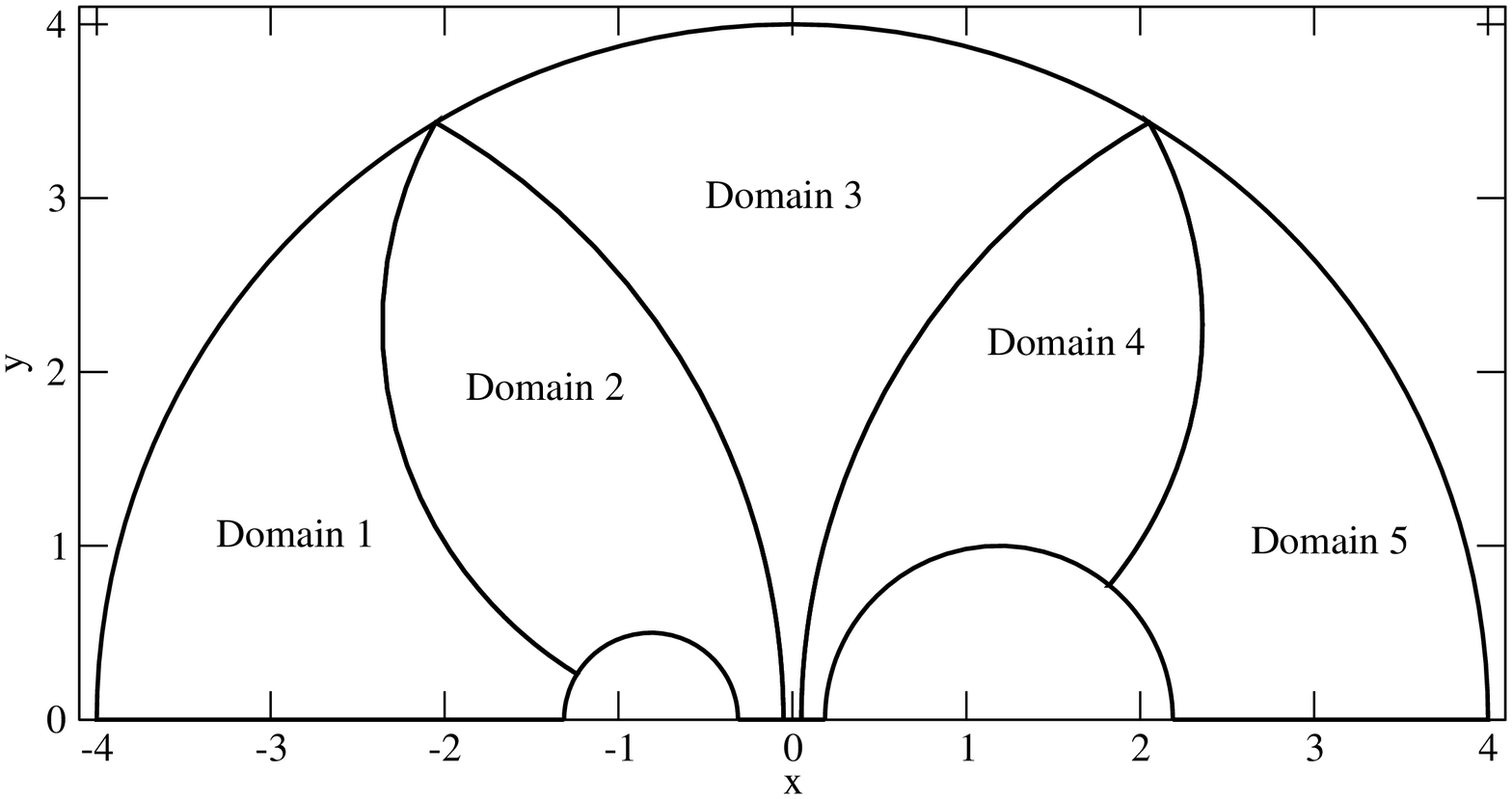} 
\caption{ \label{f:bispheric}
Multi-domain decomposition of the bispherical space. The inner spheres have radii of $0.5$ and $1$ and are separated by a distance of $2$. The radius of the outer sphere is set to $4$. The left panel shows the bispherical coordinates, along with the excision region around $\eta=0$ and $\chi=0$. The associated Cartesian coordinates, in the $\left(x, y\right)$ plane, are shown on the right panel.}
\end{figure}

\begin{table}
\caption{Relation between the various domains of the bispherical space. The table shows to what correspond the boundaries of each domain. The labels are the same as in Fig. \ref{f:bispheric}}
\begin{tabular}{c||c|c|c|c}\label{t:domains_bi}
Domain & $\eta^\star=-1$ & $\eta^\star=1$ & $\chi^\star=0$ & $\chi^\star=1$\\
\hline
1 & left inner sphere & outer sphere & $x$-axis & domain 2  \\
2 & left inner sphere & domain 3 & $x$-axis & domain 1 \\
3 & domain 2 & domain 4 & $x$-axis & outer sphere \\
4 & right inner sphere & domain 3 & $x$-axis & domain 5 \\
5 & right inner sphere & outer sphere & $x$-axis & domain 4 
\end{tabular}
\end{table}

\subsection{Cylindrical space}\label{ss:cylindrical}

Standard three-dimensional cylindrical coordinates are currently not implemented in \K. The setting exposed in this section is devoted to the study of critical phenomena in general relativity (see Sec. \ref{ss:critic} and references therein). The sought critical solution can be described on a two-dimensional space of cylindrical topology. If this setting is more specialized than spherical or bispherical coordinates, it is presented here as an illustration of the ability of \K to deal with various geometries.

The cylindrical space is described by the class {\tt Space\_critic}. The two coordinates are $x$ that ranges from $0$ to $1$ and $\tau$ that goes from $0$ to $2 \pi$. $x$ relates to the height of the cylinder and $\tau$ is the azimuthal angle. Near $x=0$ the fields of interest are either symmetric or antisymmetric. In order to take this symmetry into account, space is split into two types of domains:

\begin{itemize}
\item {\tt Domain\_critic\_inner} where $x$ lies in $\left[0, x_{\rm lim}\right]$ and simply relates to the numerical coordinate $x^\star$ by 
$x=x_{\rm lim} x^\star$. $x^\star$ covers the interval $\left[0, 1\right]$.
\item {\tt Domain\_critic\_outer} where $x$ is in $\left[x_{\rm lim}, 1\right]$. The numerical coordinate $x^\star$ relates to $x$ by an affine law $x= \displaystyle\frac{(1-x_{\rm lim})}{2} x^\star + \displaystyle\frac{(1+x_{\rm lim})}{2}$ and spans $\left[-1, 1\right]$.
\end{itemize}

In both domains, the numerical angular coordinate $\tau^\star$ coincides with $\tau$. Given the fact that the fields have some symmetries with respect to $\tau=\pi$, it is possible to restrict $\tau^\star$ to $\left[0, \pi\right[$.

\section{Spectral expansion}\label{s:spectral}
\subsection{Generalities}\label{ss:spectral}
An extensive discussion of spectral methods is beyond the scope of this work. Only the basic properties required for the comprehension of the paper are presented. Refs. \cite{GotllO77, CanutHQZ88, CanutHQZ06, CanutHQZ07, Fornb95, Boyd01} could be useful for the reader interested in more details.

In the one-dimensional case, let $f\left(x\right)$ be a function on an interval $\Lambda$. Given a set of known orthogonal functions $\Phi_i\left(x\right)$ on $\Lambda$, spectral theory enables to construct an approximation of $f$ in terms of a finite sum of the $\Phi_i\left(x\right)$. This sum is called the interpolant of $f$ and is expressed as:
\be\label{e:interpolant}
I_N f \left(x\right) = \sum_{i=0}^N a_i \Phi_i \left(x\right)
\ee
where $N$ is the order of the approximation. Typically, the basis functions are either orthogonal polynomials like Legendre or Chebyshev ones, or trigonometrical functions. In this latter case, spectral theory is nothing but the theory of discrete Fourier transform.

It can be shown that there exist $N+1$ points $x_i$ inside $\Lambda$ such that $f$ and its interpolant exactly coincide at those points 
\be
f\left(x_i\right) = I_N f \left(x_i\right).
\ee
Those points are called the collocation points. In \K, one works with the so-called Gauss-Lobato points which ensure that the boundaries of the interval $\Lambda$ coincide with the first $x_0$ and last $x_N$ collocation points.

Spectral theory provides a rule to compute the coefficients $a_i$ in terms of the values at the collocation points. Thanks to Eq. (\ref{e:interpolant}), the reverse operation, that is the computation of the $f\left(x_i\right)$ in terms of the $a_i$, is also possible. So a function $f$ can be described either in terms of its coefficients or in terms of its value at the collocation points. Depending on the operation to be performed on $f$, a description could be more useful that the other. One is working in the {\em coefficient space} when using the $a_i$ and in the {\em configuration space} when considering the $f\left(x_i\right)$. The two descriptions are completely equivalent in the sense that there is no information lost when going from one space to the other.

The most appealing feature of spectral methods is the very fast convergence of the interpolant $I_N f$ to the real function $f$. Indeed, it can be shown that, if $f$ is a \cinf function, then $I_N f$ converges to $f$ faster than any power-law of $N$. This is known as the {\em spectral convergence}. In the case of analytic functions, the convergence is even exponential. Such fast convergence enables to reach good accuracy with only moderate number of points, especially compared to other methods like finite difference schemes. As already stated in Sec. \ref{ss:multidomain} an appropriate multi-domain setting can usually ensure that the functions are \cinf in each domain even for globally less regular functions. Failing to do so will make the convergence only follow a power-law. This is called the Gibbs phenomenom in the case of a discontinuous function.

\subsection{Scalar fields}\label{ss:base_scalar}

The choice of basis functions is crucial to the success of any spectral solver. Moreover it is often a very good way of checking equations. Indeed, for complicated ones, where many terms are involved, all of them must end up having consistent basis. This section is devoted to the case of scalar fields. Higher order tensors are discussed in Sec. \ref{ss:base_higher}.

\subsubsection{Spherical coordinates}\label{sss:spher_base}

In the case of spherical coordinates, the standard basis of decomposition of a scalar field is obtained by demanding that this field can be expressed as a polynomial in terms of the Cartesian coordinates, as what is done in \cite{BonazGM99}. This condition prevents the appearance of singularities on the axis and at the origin. In addition, one requires that the fields are either symmetric or antisymmetric with respect to the plane $z=0$. This last assumption covers most of the situations of interest. By expressing the Cartesian polynomials in terms of the spherical coordinates, one can get some constraints on the basis of expansion.

Given that no symmetry is assumed with respect to the azimuthal angle $\varphi$, a standard discrete Fourier transform is performed. The basis of decomposition consists of the trigonometrical functions $\cos\left(m\varphi\right)$ and $\sin\left(m\varphi\right)$. The collocation points are the $\varphi_i = 2 \pi \displaystyle\frac{i}{N+1}$ with $0 \leq i \leq N$.

Concerning the zenith angle $\theta$, the symmetry with respect to the plane $z=0$ is taken into account. This implies that the collocation points are the 
$\theta_j = \displaystyle\frac{\pi}{2}\displaystyle\frac{j}{N}$. The associated basis functions are trigonometrical functions of one given type and of one given parity. The exact choice depends on both the function and the $\varphi$ basis. More precisely, for a symmetric function, one uses even cosines $\cos\left(2j\theta\right)$ when the basis in $\varphi$ is such that $m$ is even and odd sines $\sin\left((2j+1)\theta\right)$ otherwise. For antisymmetric functions, odd cosines are used for $m$ even and even sines otherwise.

The basis with respect to the radial numerical coordinate $r^\star$ depends on the type of domain. For both the shells and the compactified domain, it is the standard Legendre of Chebyshev polynomials with the associated Gauss-Lobato collocation points. In the nucleus, only polynomials of a given parity are considered, in order to impose some regularity at the origin. If the associated basis with respect to $\theta$ is even (resp. odd), with either sines or cosines, even (resp. odd) polynomials are used. This is true for both symmetric and antisymmetric functions. The collocation points are the Gauss-Lobato points that are located in $\left[0, 1\right]$.

Let us note that with this choice of basis, not any function is a true polynomial of the Cartesian coordinates. The choice made here is slightly less restrictive and has the advantage of being convenient and easy to handle. Additional regularity conditions are discussed in Sec. \ref{ss:regularity}

\subsubsection{Bispherical coordinates}\label{sss:bispheric_base}

The same guidelines as in the spherical case are used to derive appropriate basis of decomposition. One demands that scalar fields can be expressed as polynomials in terms of Cartesian coordinates. Once again, only symmetric or antisymmetric functions with respect to the plane $z=0$ are considered. This gives the following constraints on the basis.

The angle $\varphi$ is expanded on cosines $\cos\left(m\varphi\right)$ (resp. sines) for symmetric (resp. antisymmetric) functions. The associated collocation points are the $\varphi_i = \pi \displaystyle\frac{i}{N}$.

For the coordinate $\chi^\star$, due to regularity conditions on the $x$-axis, only polynomials (Chebyshev or Legendre) of some given parity must be taken into account. The parity depends on the basis with respect to $\varphi$. Even polynomials must be used when $m$ is even and odd otherwise. This can be seen in the expressions of the Cartesian coordinates given by Eqs. (\ref{e:bispheric}). This is true for both symmetric and antisymmetric functions. The collocation points are the Gauss-Lobato points that are located in $\left[0, 1\right]$.

The coordinate $\eta^\star$ is expanded on standard polynomials, either Chebyshev or Legendre. The collocation points are the usual Gauss-Lobato points and span the interval $\left[-1, 1\right]$.

As in the spherical case, this choice of basis is slightly less restrictive than what would be needed to get true Cartesian polynomials (see Sec. \ref{ss:regularity} for more regularity conditions).

\subsubsection{Cylindrical coordinates}\label{sss:cylindrical_base}

As already stated, the functions considered in the critical phenomenom case (see Sec. \ref{ss:critic}) are either symmetric or antisymmetric with respect to $x=0$. To account for this fact, in the inner domain, only polynomials of the appropriate parity are used, either Chebyshev or Legendre. The collocation points in terms of $x^\star$ are the Gauss-Lobato points and are located in $\left[0, 1\right]$. In the outer domain, standard polynomials are used.

Some symmetry is also taken into account with respect to $\tau=\pi$. If a function $f$ is such that $f\left(\tau+\pi\right)=f\left(\tau\right)$ then only even trigonometrical functions are considered and $f$ is expanded on both $\cos\left(2 i \tau\right)$ and $\sin\left(2 i \tau\right)$. The other possibility is a function $g$ such that $g\left(\tau+\pi\right)=-g\left(\tau\right)$. In that case, only odd trigonometrical functions are used.

\subsection{Higher rank tensors}\label{ss:base_higher}

For tensors, the choice of spectral basis depends on both the geometry and the tensorial basis. The simplest choice is the one of a Cartesian tensorial basis. Let us precise that this choice can be made almost independently of the geometry, as long as Cartesian coordinates are defined. For instance, a vector $\vec{V}$ can be described by its Cartesian components $\left(V^x, V^y, V^z\right)$ in spherical geometry, meaning that each component is given in terms of $\left(r, \theta,\varphi\right)$.

Let us first turn to the case of a vector with a Cartesian tensorial basis. The appropriate basis are obtained by demanding that this vector can be expressed as the gradient of a scalar field. Given this assumption, it simply follows the each component can be expressed as a polynomial of the Cartesian coordinates. Thus each component behaves like a scalar field and the same spectral basis are used. As far as the symmetry $z=0$ is concerned (for the spherical and bispherical cases), it is assumed that the components $x$ and $y$ of the vector are symmetric and the $z$ one antisymmetric (i.e. the vector is the gradient of a symmetric scalar field). The same is true for higher order tensors in Cartesian tensorial coordinates for they can be obtained as tensorial product of vectors. When needed, the appropriate $z=0$ symmetry is also obtained from this fact.

The case of a spherical orthonormal tensorial basis is also implemented, when one is working in spherical geometry. The appropriate spectral basis of each component can be derived from the Cartesian case by making a careful use of the passage formulae that relates the two basis. For instance, if the radial component of a vector behaves like a (symmetric) scalar, this is not the case of the $\theta$ one for which the spectral basis with respect to $\theta$ involves even sines for $m$ even and odd cosines otherwise, as can be seen from the formula
\be
\label{e:passage}
V^\theta = V^x \cos \theta \cos \varphi + V^y \cos \theta \sin \varphi - V^z \sin \theta.
\ee

In the current state of \K, only Cartesian tensorial basis are implemented in the bispherical case. In the spherical case, both Cartesian and orthonormal spherical tensorial basis are defined. The cylindric space only deals with scalars so far. Additional cases will be implemented when needed.

\subsection{Additional regularity conditions}\label{ss:regularity}
\subsubsection{Galerkin basis}
As stated in Sec. \ref{ss:base_scalar}, the spectral basis chosen do not ensure a complete regularity of the fields. Some additional constraints must be enforced by means of an appropriate Galerkin technique which goes as follows. The fields are not expanded onto any set of basis functions but only onto a subspace, which verifies the additional constraints one wishes to enforce. For instance, let us consider a one-dimensional function $f\left(x\right)$ expanded onto even Chebyshev polynomials $T_{2i}\left(x\right)$. To fulfill the additional constraint that $f$ vanishes at $x=0$ it is possible to use the Galerkin basis of the $G_i = T_{2i+2}+ \left(-1\right)^{i+1}$. $f$ is expanded onto the $G_i$ and thus, by construction, does verify the constraint $f\left(0\right)=0$. Let us note that, in general, a Galerkin basis is not orthogonal.

\subsubsection{Regularity on the context of spectral methods}\label{sss:spectral_regularity}
Given the geometries present in \K, two types of regularity must be discussed. The regularity on one axis that must be enforced in both spherical and bispherical cases and the regularity at the origin $r=0$ in the case of spherical coordinates only. The main reason why those regularities must be carefully handled can be found in the way spectral methods compute some ratios.

As an illustration, let us concentrate on the axis case in spherical coordinates. For many operators, some functions must be divided by $\sin\theta$. This is the case of the Laplacian of a scalar $f$ where terms like $ \displaystyle\frac{\cos \theta}{r^2 \sin\theta}\displaystyle\frac{\partial f}{\partial\theta}$ do appear. The division by $\sin\theta$ can be troublesome on the axis where it vanishes. In the case of spectral methods this difficulty can be overcome by working in the coefficient space thus using the fact that the spectral approximation is not local. However, doing so, what is computed is not the exact ratio of a function $g$ by $\sin\theta$ (that would diverge for arbitrary function) but the regularized one $R = \displaystyle\frac{g - g\left(\theta=0\right)}{\sin\theta}$. $R$ is obviously regular for any function $g$ because the finite part of $g$ on the axis has been taken out. It does coincide with the real ratio if and only if $g$ vanishes on the axis. 

There is another way to look at this problem. When one computes the Laplacian of a function in spectral method, what is actually computed is not the real Laplacian, but another operator that includes the finite parts on the axis. Those finite parts cause the appearance of additional homogeneous solutions (typically solutions that do not vanish on the axis). Some examples of that can be found in \cite{GrandBGM01}, for the radial coordinate, either at the origin or at infinity. The extra homogeneous solutions must be dealt with by enforcing additional conditions on the solution. If one fails to do so, the resulting system will not be invertible. The additional conditions are what we refer as regularity conditions and are discussed in more detail in Sec. \ref{sss:axis_regularity} and \ref{sss:origin_regularity}.

\subsubsection{Regularity on the axis}\label{sss:axis_regularity}

Let us first consider the case of a scalar field $f$ in spherical coordinates. The regularity requires that $f$ vanishes on the axis, except for the $m=0$ case. Given the basis of decomposition discussed in Sec. \ref{sss:spher_base} this can be enforced by using the Galerkin basis $\cos\left(\ell \theta\right) -1$ instead of standard cosines, when $m \not= 0$. When the basis with respect to $\theta$ involves sines only, there is no need to impose any additional condition. In the case $m=0$ the standard basis are used.

For higher order tensors the same guidelines as in Sec. \ref{sss:spher_base} are used. When a Cartesian tensorial basis of decomposition is employed, the regularity conditions for each component are the same as in the scalar case. With a spherical tensorial basis, they are slightly different. Typically one can allow some components (the angular ones) to take non-zero values on the axis for higher $m$. For instance, the component $V^\theta$ of a vector $\vec{V}$ relates to the Cartesian components by the passage formula (\ref{e:passage}). Given the regularity conditions for the Cartesian components, it is easy to see that $V^\theta$ can be non-zero on the axis for $m=1$.

In the bispherical case, the situation is very similar. As seen in Sec. \ref{ss:bispheric}, the axis is described by $\chi^\star=0$. For scalars or Cartesian components of tensors, one demands that they vanish on the axis, for $m>0$. When even Chebyshev (resp. Legendre) polynomials are used, this is done by using the Galerkin basis: $T_{2i}\left(\chi^\star\right) + \left(-1\right)^{i+1}$ (resp. $L_{2i}\left(\chi^\star\right) - L_{2i}\left(0\right)$). When odd functions are used, not additional condition is enforced.

\subsubsection{Regularity at the origin}\label{sss:origin_regularity}

The situation at the origin in the case of a spherical geometry is more complicated. It can be shown that a true polynomial of Cartesian coordinates would vanish as $r^\ell$ at the origin, where the $\theta$ basis is $\cos\left(\ell \theta\right)$ or $\sin\left(\ell \theta\right)$ (see \cite{BonazGM99}). However this condition is difficult to enforce exactly and, in some cases, has been found to generate instabilities.

As a first step one demands that for $\ell \not=0$ the functions vanish at the origin. When using even Chebyshev (resp. Legendre) polynomials in $r$, this is done by using the Galerkin basis: $T_{2i+2}\left(r^\star\right) + \left(-1\right)^{i+1}$ (resp. $L_{2i}\left(r^\star\right) - L_{2i}\left(0\right)$). Nothing needs to be done if odd polynomials are used. This choice of basis enforces the first order of the regularity at the origin. 

However, this is not sufficient to ensure that usual second order operators like the Laplacian are well inverted. This is related to the appearance of additional homogeneous solutions discussed in Sec. \ref{sss:spectral_regularity}. The second order of the regularity condition must be supplemented. This is done by demanding that the derivative with respect to $r$ vanishes at the origin, for $\ell>1$. If this is automatically verified for even polynomials, this is not the case for odd ones. When dealing with odd Chebyshev polynomials a possible choice of Galerkin basis is given by 
$T_{2i+3} - T'_{2i+3}\left(0\right) T_1$, for $\ell>1$ (and similarly for Legendre).

This choice has proven to be sufficient to ensure that second order differential equation are solved properly at the origin. For higher order equations, there may be need to further strengthen regularity. Let us note that, contrary to the regularity on the axis, the regularity conditions at the origin are the same for scalars or components of tensors. This comes from the fact that the transformations that relate the Cartesian components to the spherical ones do not involve $r$ (see for instance (\ref{e:passage})).

The regularity on the axis must obviously also be enforced when the origin is present. This is resulting in a two dimensional Galerkin basis for $r$ and $\theta$. Putting all the pieces together, the appropriate basis of decomposition for a scalar field symmetric with respect to $z=0$, in a spherical nucleus would be

\begin{itemize}
\item For $\varphi$ : the standard Fourier decomposition in $\cos \left(m \varphi\right)$ and  $\sin \left(m \varphi\right)$
\item For $\theta$ (regularity on the axis): \begin{itemize}
		     \item $\cos \left(2j \theta\right)$ for $m=0$ ($\ell = 2j$)
		     \item $\cos \left(2j \theta\right) -1$ for $m$ even and $m \not= 0$ ($\ell=2j$)
		     \item $\sin \left(\left(2j+1\right) \theta\right)$ for $m$ odd ($\ell = 2j+1$)
		      \end{itemize}
\item For $r$ (regularity at the origin): \begin{itemize}
		     \item $T_{2i}\left(r^\star\right)$ for $\ell=0$
		     \item $T_{2i+1}\left(r^\star\right)$ for $\ell=1$
		     \item $T_{2i+2}\left(r^\star\right) + \left(-1\right)^{i+1}$ for $\ell$ even and $\ell \not= 0$
		     \item $T_{2i+3}\left(r^\star\right) - T'_{2i+3}\left(0\right) T_1$ for $\ell$ odd and $\ell \not= 1$.
		     \end{itemize}
\end{itemize}

Let us mention that if the regularity on the axis is routinely handled in the context of spectral methods, this is not the case of the conditions at the origin. For many applications, like the meteorology simulations  or the black holes with excision (see Sec. \ref{ss:hkv} and \ref{ss:kerr}), the origin is not part of the computational domain. Even when the origin is present, various tricks are usually employed to avoid dealing with regularity conditions. For instance, in \cite{FoucaKPT08, Tichy09} the region close to the origin is not described by spherical coordinates but by Cartesian ones. In \cite{AnsorBT04}, the solution is sought in the configuration space with collocation points that avoid the origin.

\subsection{Effective number of unknowns}\label{ss:unknowns}

\K is designed to solve systems of equations in the coefficient space. Using such a setting the unknowns are the true coefficients of the fields of interest. By true coefficients one means that some of them are irrelevant. For instance when using a standard Fourier transform the coefficient of $\sin\left(0\right)$ is obviously meaningless. Some cases are less trivial. Indeed, the last coefficient of an expansion in terms of odd Chebyshev polynomials is also irrelevant. This is related to the fact that the value of such a function is, by construction, 0 at $x=0$. The value of the function is freely specifiable only at $N-1$ of the $N$ collocation points. To maintain the bijection between the coefficient space and the configuration space one then needs to reduce the number of true coefficients.

Moreover the regularity conditions discussed in Sec. \ref{ss:regularity} reduce furthermore the true number of unknowns. As already stated, such additional conditions are imposed by a Galerkin method. This implies that the unknowns are not the coefficients onto the standard basis of expansion but rather onto the Galerkin basis. The number of coefficients is then different, reflecting the fact that the Galerkin basis contains informations about additional conditions. For instance, suppose one works with $N+1$ coefficients in terms of even Chebyshev polynomials so that the function is expanded onto the $T_{2i}$ with $0 \leq i \leq N$. Suppose now that the Galerkin basis $T_{2i+2}\left(r^\star\right) + \left(-1\right)^{i+1}$ is used to enforce that the function vanishes at the origin (like in Sec. \ref{sss:origin_regularity}). It is easy to see that, in order to keep the same degree of approximation (i.e. to truncate the series to the same order in terms of the polynomials), one needs to go only up to $i=N-1$. This is a general feature of the Galerkin method.

The true number of coefficients and so the number of true unknowns is determined by \K by making use of the type of field considered (scalar, vector, rank-2 symmetric tensor...), the symmetry, the tensorial basis and the geometry. Such computation is automated and should be transparent to the user.

\section{Setting equations}\label{s:equations}
\subsection{The weighted residual method}\label{ss:weighted}

The usual way of solving partial differential equations in the context of spectral methods is based on the framework of the weighted residual methods (see for instance \cite{GrandN09} for a more detailed presentation). Let us consider an equation written formally as $R=0$. $R$ is a general function on the space of interest. For instance, if one needs to solve a simple Poisson equation, one would have $R= \Delta f - S$, where $S$ is the source and $f$ the unknown field. Obviously, more complicated, non-linear cases involving several unknown fields are possible. The weighted residual method translates the functional equation $R=0$ into a finite set of discrete equations by demanding that the scalar product $\left(R, \xi\right)$ of $R$ with respect to some test function $\xi$ vanishes. The scalar product is the same as the one used to define the spectral expansion.

Depending on the choice of test functions one generates different methods. \K mainly implements the variant known as the $\tau$-method. In this case the test functions are the same as the spectral basis ones. Suppose that $R$ is expanded onto the $\xi$ by $R=\sum C\left(\xi\right) \xi$, $C$ being the coefficient corresponding to the basis function $\xi$. The sum is taken on all the dimensions, up to the desired order of approximation. One can then show that the residual equations $\left(R, \xi\right) = 0$ is equivalent to demanding that $C\left(\xi\right)=0$. Doing so, one then obtains as many equations as the total number of basis functions. In the $\tau$-method, the equations corresponding the last coefficients can be relaxed and replaced by equations that describe appropriate matching between the domains and boundary conditions. The exact number of equations that must be relaxed depends on both the geometry and the equations themselves.

As mentioned is Sec. \ref{ss:regularity}, when regularity conditions must be enforced, a variant of the $\tau$-method known as the Galerkin method is used. In this framework, the fields are expanded onto the Galerkin functions $G$. However the residuals $R$ are expanded onto the standard basis. In the Galerkin method, the test functions are the Galerkin ones. One then gets as many equations as the number of Galerkin functions, formally written as $\left(R, G\right)=0$. One can show that such equations are linear combinations of the coefficients of $R$ which depend on the Galerkin basis used. As for the $\tau$-method, some equations can be relaxed to enforce additional conditions.

To summarize, one can say that \K solves equations in the coefficient space. Standard $\tau$-method is used to impose appropriate matching and boundary conditions. When regularity conditions must be enforced, this is done by means of a Galerkin method.

\subsection{Equations for the fields}\label{ss:fields_eqs}

The discretization of the field equations by means of the $\tau$ and Galerkin methods are implemented by the class {\tt Equation} and its derived classes. Given the knowledge of the geometry and the type of equation, those objects are able to produce the appropriate number of discrete residual equations.

The most widely used derived class is called {\tt Eq\_inside}. It deals with equations that must be solved inside a given domain, for instance a Poisson equation of the type $\Delta f - S =0$. For each type of domain, the $\tau$ or Galerkin residual equations can be generated. Depending on the number of boundaries, several equations are relaxed in order to impose matching and boundary conditions. By default {\tt Eq\_inside} assumes that the equation to be solved contains second order derivative of the fields. For first order equations (resp. zeroth order) one would use the related class {\tt Eq\_one\_side} (resp. {\tt Eq\_full}).

Boundary conditions are encoded in the class {\tt Eq\_bc} which must be supplied with the domain and the type of boundary. $\tau$ or Galerkin method of appropriate dimensions are constructed by this class. For instance, the boundary conditions on a sphere are written with a two-dimensional Galerkin method with respect to the angles $\left(\theta, \varphi\right)$, in order to account for the regularity conditions on the axis.

The class {\tt Eq\_matching} is used to impose the matching of quantities across the boundary between two domains. This is done in the coefficient space and so must be used only when the basis relative to the surface is the same on both side of the boundary. This may seem restrictive but it covers most of the geometries implemented in \K, at least when the same resolution is used in each domain. This is the case for all the boundaries of the spherical geometry but not for the boundary between the bispherical domains and the spherical compactified one (see Sec. \ref{ss:bispheric}). In most of the cases the quantity to be matched is simply an unknown field or its normal derivative. Nevertheless it is possible to impose the matching of different quantities across the boundary. This would be useful in the case where different variables are used in different domains (see an example in Sec. \ref{ss:critic}).

When the basis are not the same across the boundaries, because of different geometries or different resolutions, one must use the class {\tt Eq\_matching\_non\_std} that performs the matching in the configuration space. In this case, given the fact that the collocation points are usually different on the two sides of the boundary, one must choose the set of points at which the matching conditions are written. It is up to the user to make sure that this choice is consistent and leads to a global problem that is not over or under-determined.

\subsection{Integral equations}\label{ss:integral_eqs}

\K also enables to impose global conditions on the fields. One could prescribe the value of the total energy content of space or the integral of a given field on some surface, for instance. If the associated equations are computed in terms of the coefficients of the fields, they are not functional equations and do not require the machinery of the weighted residual method. They simply translate into additional conditions that must be fulfilled by the fields.

In order to ensure that the full system contains the same number of unknowns than conditions, integral equations are generally associated with global unknowns. This is for instance the case of black holes, where the local rotation rate is an unknown that is constrained by demanding that a global quantity, the spin, takes a given value (see Sec. \ref{ss:hkv}). Another example is found in the critic solution case of Sec. \ref{ss:critic}, where the period of the solution is sought by imposing that one of the modes of one of the fields vanishes at the center.

\section{Solving the system}\label{s:system}
\subsection{Newton-Raphson iteration}\label{ss:newton}

It follows from the techniques described in Sec. \ref{s:spectral} and \ref{s:equations}, that the solver translates a system of functional equations into a finite set of algebraic ones. In general, those equations are non-linear and \K relies on the well-known Newton-Raphson technique to find a solution (as in \cite{PfeifKST03, AnsorKM03}). Let us consider a set of unknowns, formally denoted by the vector $\vec{u}$. The system of equation can be written as $\vec{f} \left(\vec{u}\right)=0$. Obviously the vectors $\vec{u}$ and $\vec{f}$ must have the same size.

The solution is sought by iteration, starting from an initial guess $\vec{u}^0$. Let us denote $\vec{u}^n$ the approximation of the solution found after n$^{\rm th}$ iterations. The Newton-Raphson scheme proceeds as follows. First one computes the vector $\vec{f}^n = \vec{f}\left(\vec{u}^n\right)$. If $\vec{f}^n$ is small in some appropriate sense (for instance the maximum norm is smaller than a given threshold), then $\vec{u}^n$ is good enough solution. If not, one needs to compute the Jacobian matrix ${\bf J}^n$ of the system defined as:
\be
J^n_{ij} = \frac{\partial f_i}{\partial u_j} \left(\vec{u}^n\right)
\ee
where $f_i$ and $u_j$ denote to the ith and jth} components of $\vec{f}$ and $\vec{u}$ respectively. One row of the Jacobian corresponds to one equation, and one column to the derivation with respect to one of the unknowns. The matrix is computed at the position of the current solution $\vec{u}^n$ and so must be recalculated at each iteration. 

The linear system 
\be\label{e:linear}
{\bf J}^n \vec{X}^n = \vec{f}^n
\ee
must then be solved and the next approximation of the solution is given by $\vec{u}^{n+1} = \vec{u}^n - \vec{X}^n$. The method is known to converge rapidly to the solution, at least if the initial guess is good enough. For linear systems, the Newton-Raphson method finds the solution in one iteration step.

\subsection{Automatic differentiation}\label{ss:automatic}

For simple problems it can be possible to explicitly derive an analytic expression of the Jacobian matrix from the knowledge of the equations $\vec{f}$. This can be done by making use of some software that can do symbolic computations. Such analytic expression can then be fed to the code and used to compute the Jacobian. However, in practice, that can be tricky when the equations are complicated, especially if their explicit expressions in terms of the unknowns are intricate (this is especially true for Sec. \ref{ss:kerr}, where an explicit expression of the equations in terms of the spatial metric would be almost impracticable).

In the context of \K, the computation of the Jacobian is done numerically by a technique of automatic differentiation based on the notion of dual numbers. Each quantity $x$ is supplemented by its infinitesimal variation $\delta x$. The new object is usually denoted as 
$\left<x, \delta x\right>$. The arithmetic of those dual objects is then implemented in order to accommodate the usual rules for differentiation. For instance one gets:
\bea
\left<x, \delta x\right> + \left<y, \delta y\right> &=& \left<x+y, \delta x+ \delta y\right> \\
\left<x, \delta x\right> \times \left<y, \delta y\right> &=& \left<x y,  x  \delta y + y \delta x\right> \\
\sqrt{\left<x, \delta x\right>} &=& \left< \sqrt{x}, \frac{\delta x}{2\sqrt{x}}\right>.
\eea

Let us consider a set of values $\vec{u}$ of our unknowns and let us supplemented it with a set of infinitesimal variations $\delta \vec{u}$. Using the extended algebra one can apply  the system of equations $\vec{f}$ to $\left< \vec{u}, \delta\vec{u}\right>$. This would result in the following extended object $\left< \vec{f}\left(\vec{u}\right), \delta \vec{f} \left(\vec{u}, \delta\vec{u}\right)\right>$. The first part 
$ \vec{f}\left(\vec{u}\right)$ is the usual application of the system of equations to $\vec{u}$ and would be used to get the term $\vec{f}^n$ in the Newton-Raphson iteration (see Sec. \ref{ss:newton}).

One can show that the second part $\delta \vec{f}$ is a vector that is the product of the Jacobian with the vector $\delta\vec{u}$ so that:
\be
\delta \vec{f} \left(\vec{u}, \delta\vec{u}\right) = \left({\bf J}\left(\vec{u}\right)\right) \times \delta \vec{u}.
\ee
The use of the dual numbers enables to automatically compute the product of the Jacobian with any vector. In order to compute the whole Jacobian matrix, one then proceeds as follows. First the unknowns are supplemented with a variation vector that is zero except at the $i^{\rm th}$ component which is set to one. The system of equation is applied, in its extended form, to this object. The variation of the result is then, by construction, 
the $i^{\rm th}$ column of the Jacobian. By taking all the possible values of $i$, the whole matrix can be generated.

\subsection{Parallelization and linear system solvers}\label{ss:parallel}

The size of the linear system (\ref{e:linear}) is the same as the total number of unknowns. If one considers $N_f$ scalar fields, in $N_d$ domains of $d$ dimensions and if $N$ is the number of coefficients of the spectral expansion in each dimension, then the Jacobian is an $\left(m \times m\right)$ matrix with $m \approx N_f N_d N^d$. This number can be quite large, especially in three dimensions. For instance, for $N_f=5$, $N_d=6$, $d=3$ and $N=21$, which are big but not huge numbers, one would get $m \gtrsim 250,000$. The resulting matrix would represent more than 500 GB of data. Such amount of data would be difficult to store on a single processor, without mentioning the number of operations required for the inversion.

There are several ways to overcome the difficulty of dealing with such big matrices. One solution is to use an iterative technique to solve the system (\ref{e:linear}) (see \cite{Saad03} for a general introduction to the iterative techniques and  \cite{PfeifKST03,Ansor05} for some applications). Doing so, the solution of the linear system is sought in a loop that does not require the storage of the full matrix ${\bf J}$. It is only needed to be able to compute the product ${\bf J} \vec{x}$ for any vector $x$. If this is exactly the information that is available from our automatic differentiation technique (see Sec. \ref{ss:automatic}), \K, in its current state, does not use this kind of algorithms. Indeed, the tests conducted with the iterative solvers, in the context of \K, showed a lack of stability, meaning that the convergence was achieved only for some problems and only for moderate degrees of freedom. The reason for that must probably be sought in the preconditioning step. Indeed, the success of the iterative algorithms relies on the knowledge of an efficient preconditioning matrix that approximates ${\bf J}^{-1}$. If some preconditioning techniques have been successfully used in the context of spectral methods like in \cite{PfeifKST03, Ansor05}, they are only applicable when working in the configuration space, which is not the case of \K. Nevertheless, for very large problems, it is likely that the use of iterative techniques would be highly desirable. This is why their implementations will be an important axis of future development of the library.

That being said, in its current state, \K relies on a direct method to invert the linear system. In order to deal with very large matrices, distributed algorithms are available. Typically, each processor has the knowledge of only some parts of the matrix. In this context, a parallelization of \K is straightforward and is done via MPI. Indeed, the Jacobian is computed columns by columns, each computation being independent of the other. Each processor computes and stores a manageable set of columns. The size of the Jacobian is only limited by the total amount of memory available. This part of the computation is perfectly linear in terms of the number of processors.

After having tested several libraries, it appeared that a parallel version of {\tt LAPACK} called {\tt Scalapack} \cite{scalapack}, was the best suited. However, before calling the function that computes the solution the linear system, a redistribution of the matrix amongst the various processors must be done. The reason for this is purely computational and its goal is to ensure a better behavior of {\tt Scalapack} algorithms. Let us precise that {\tt Scalapack} library does provide the function needed to do this redistribution. The LU decomposition of ${\bf J}$ itself does not require more memory than what is needed to distribute the Jacobian across the various processors. Concerning the computational time, it has been observed that the resolution of the linear system takes roughly as much time as what is needed to compute the Jacobian itself, at least for the cases exhibited in Sec. \ref{s:test}.

For small problems (typically 2-dimensional problems in low resolution), a sequential version of \K can be used, where the linear system is solved using the standard {\tt LAPACK} library \cite{lapack}.

\section{User interface}\label{s:user}

If \K is a rather intricate tool, a great effort has been made in its design to render its use as simple as possible. This is especially true in the way the equations are passed to the solver. In this section the various steps in which \K should usually be called are briefly summarized. \K is written in C++ and it is probably best if the user has at least some knowledge of this language. What follows does not constitute an extensive documentation of the library but is intended to give a feeling of what a typical use of \K is. For more details, the would-be user is strongly advise to take a look at the codes that have be made available in \K repository (this is the case of the four examples discussed in Sec. \ref{s:test}).

The first step is to specify the geometry of space. This is done by calling the constructor of one of the derived classes of {\tt Space} (see Sec. \ref{s:geometry}). Various parameters describing the geometry are required at this point, like the number of shells for a spherical space, or the various radii involved in a bispherical one. The constructor must also be supplied with the required resolution (i.e. the number of points) and the type of polynomials to be used (Legendre or Chebyshev).

On the desired geometry, the user needs to define the fields of interest. They can be of various type, scalars, vectors, higher order tensors, metric tensors, etc. As the solution is sought by iteration, those objects are usually set to some initial guess. The choice of the initial values depends very much on the problem. It can come from reading data from another code or by finding appropriate analytic expressions. A crucial point of this step is to affect each field with its correct spectral basis of decomposition. For most cases, \K provides functions that do this automatically but the user should make sure that those functions are appropriate for each given problem. Failing to provide the right basis will either cause the solver to abort (if it needs to sum two quantities with incompatible basis for instance), or cause the appearance of Gibbs-like phenomenom (if a general function is expanded onto symmetric basis for instance).

All the informations needed to solve a system are contained in the class {\tt System\_of\_eqs}. It is constructed from the space of interest and can be supplied with the domains on which the system is to be solved, in case the values of fields are not defined on the whole space. The list of variables and constants are then passed to the system. By variables one means quantities that are unknowns and that must be solved for, whereas constants are quantities that can appear in the equations but have fixed values. Both type of objects can be either fields or numbers. Each variable and constant must be a quantity that has been initialized beforehand. Each of them is supplied with a character string that is the name by which it will by recognized in the equations. A special field is the metric one, when defined, for which additional functionalities are available (tensorial indices manipulation, covariant derivative, etc.)

The equations are passed to the {\tt System\_of\_eqs} as character strings. Those strings are read by \K to generate the appropriate computation rules. The formalism used is inspired by LateX \cite{latex}, especially in the way indices are handled. Einstein's convention of summation on repeated indices is used (for instance, in the 3-dimensional Cartesian case, $f_i * g^i = f_x g^x + f_y g^y + f_z g^z$). Various reserved words are available to encode some functions like $\tt sqrt$ that stands for the square root, $\tt dn$ for the normal derivative with respect to one surface, or $\tt R$ for the Ricci scalar of a given metric. When a quantity appears many times in the equations, it can be made into a definition. Apart from the fact that it can simplify the writing of the equations, it has the advantage that definitions are computed only when the value of the variables change, and not each time they are encountered, thus resulting in a faster computation. When setting the equations, the user must not only make sure they are correct but also that they give raise to as many conditions as the number of unknowns, in order to get an invertible system. This can be tricky at times, especially when different resolutions are used in different regions of space. In some standard cases, member functions of the {\tt Space} class provide means to implement equations in the whole space instead of passing them domain by domain.

Finally a solution of the system is sought, calling the member function {\tt do\_newton}, the required accuracy being passed as a parameter. At this point \K verifies that the number of unknowns is consistent with the number of equations before entering the Newton-Raphson loop. Once convergence is achieved, the various data can be saved, manipulated, results printed, etc.

\section{Test problems}\label{s:test}
In this section, four different problems are presented. The main goal is to illustrate the ability of \K to deal with different and complicated situations. The examples have been chosen to show various and different aspects of \K. Nevertheless, they are far from being merely test problems. They all related to very contemporary physics. That being said, this paper is not the place to extensively discuss physics and so explanations about this are kept to the minimal level required for comprehension. Given the current limitations of \K, the four problems have in common the fact that they all are boundary value problems on a fixed and known geometry. The results shown below were obtained by using Chebyshev polynomials. The results were roughly the same with Legendre polynomials, even if the observed convergence seems to be slightly slower in this case.

Computations were conducted on a cluster of 10 quadcore bi-opterons at 2.5 GHz. Each node has 16 Go of RAM and the cluster runs under Linux. Under those conditions, the biggest Jacobian invertible has a size just below $100,000$ and it takes 4 h to do the inversion. Most of the codes used to compute the following results are available from the {\tt Codes/Par\_version} repository of \K.

\subsection{Vortons}\label{ss:vortons}

In classical field theory, some closed vortex loops can exist. Such loops are stabilized either by twisting them, in which case one talks about knots, or by centrifugal force when the vortex is spinning. This last case is called a vorton (see \cite{RaduV08} for a review on this type of objects). In this section, vortons are computed in the so-called Wittens theory \cite{Witte85}. In this model, superconducting cosmic strings are known to exist. They are, basically, straight configuration of the fields. The idea behind vortons is to try to cut a piece of a string, make it into a loop and try to stabilize it by giving it a spin. If this idea dates back to \cite{DavisS88a, DavisS88b}, it is only recently that the such solutions were explicitly computed in \cite{RaduV08}.

From a mathematical point of view, vortons are described by two complex scalar fields $\phi$ and $\sigma$. The vorton is computed by demanding that $\sigma$ takes the particular form \be\label{e:ansatz}
\sigma = Z \exp\left[i\left(m \varphi + \omega t\right)\right]
\ee
where $Z$ is a real function, $m$ is the azimuthal winding number that constraints the topology of the vorton, $\varphi$ is the angle around the axis of the loop and $\omega$ is the angular velocity. No particular ansatz is assumed for the other field that is described by its real and imaginary parts:
$\phi = X + i Y$.

The geometry of the solution is such that the fields are axisymmetric. Moreover, $X$ and $Z$ are symmetric with respect to the plane $z=0$ and $Y$ is antisymmetric. Let us note that from the point of view of \K, $Z$ is not a scalar field but rather the harmonic $m$ of a scalar field (i.e. $Z \cos\left(m\varphi\right)$ is the real scalar field). This has to be taken into account when setting the appropriate spectral basis. This also changes the regularity conditions on the axis, where $Z$ must vanish, which would not be the case for a real scalar field (see Sec. \ref{ss:regularity} for more details). Given the geometry, the problem is defined on a polar space by the class {\tt Space\_polar} (see Sec. \ref{ss:spheric}).

Under those assumptions, the three unknown fields obey a set of three elliptic equations:
\bea
\Delta X &=& \left(\frac{\lambda_\phi}{2}\left(X^2 + Y^2 -1\right) + \gamma Z^2\right) X \\
\Delta Y &=& \left(\frac{\lambda_\phi}{2}\left(X^2 + Y^2 -1\right) + \gamma Z^2\right) Y \\
\Delta_m Z &=& \left(\frac{\lambda_\sigma}{2} \left(Z^2- \eta^2_\sigma\right) + \gamma \left(X^2+Y^2\right) - \omega^2\right) Z
\eea
where $\Delta_m$ is the Laplacian of the $m$th-component of a scalar field, so that $\Delta_m Z = \Delta Z - \displaystyle\frac{m^2 Z}{r^2\sin^2\theta}$. This expression shows that $Z$ must vanish on the z-axis. At spatial infinity the fields are such that $X=1$, $Y=0$ and $Z=0$.

The equations are passed as such to \K. Being second order differential equations, matching of the fields and of their normal derivatives is performed at the boundaries between the various domains. The main difficulty in finding solutions lies in the fact that the system involves many free parameters $\lambda_\phi$, $\lambda_\sigma$, $\gamma$, $\eta_\sigma$, $\omega$ and $m$. It is believed that solutions only exist in some parts of this huge parameter space. Another difficulty is related to the existence of trivial solutions (like $X=1$, $Y=0$ and $Z=0$ in the whole space). If one does not start the solver from a good enough initial guess, then the interesting solutions will be missed. For those reasons, it is quite an achievement that the authors of \cite{RaduV08} did identify an appropriate region of the parameter space and were able to compute the associated vortons.

\begin{figure}[htb]
\includegraphics[width=0.55\textwidth]{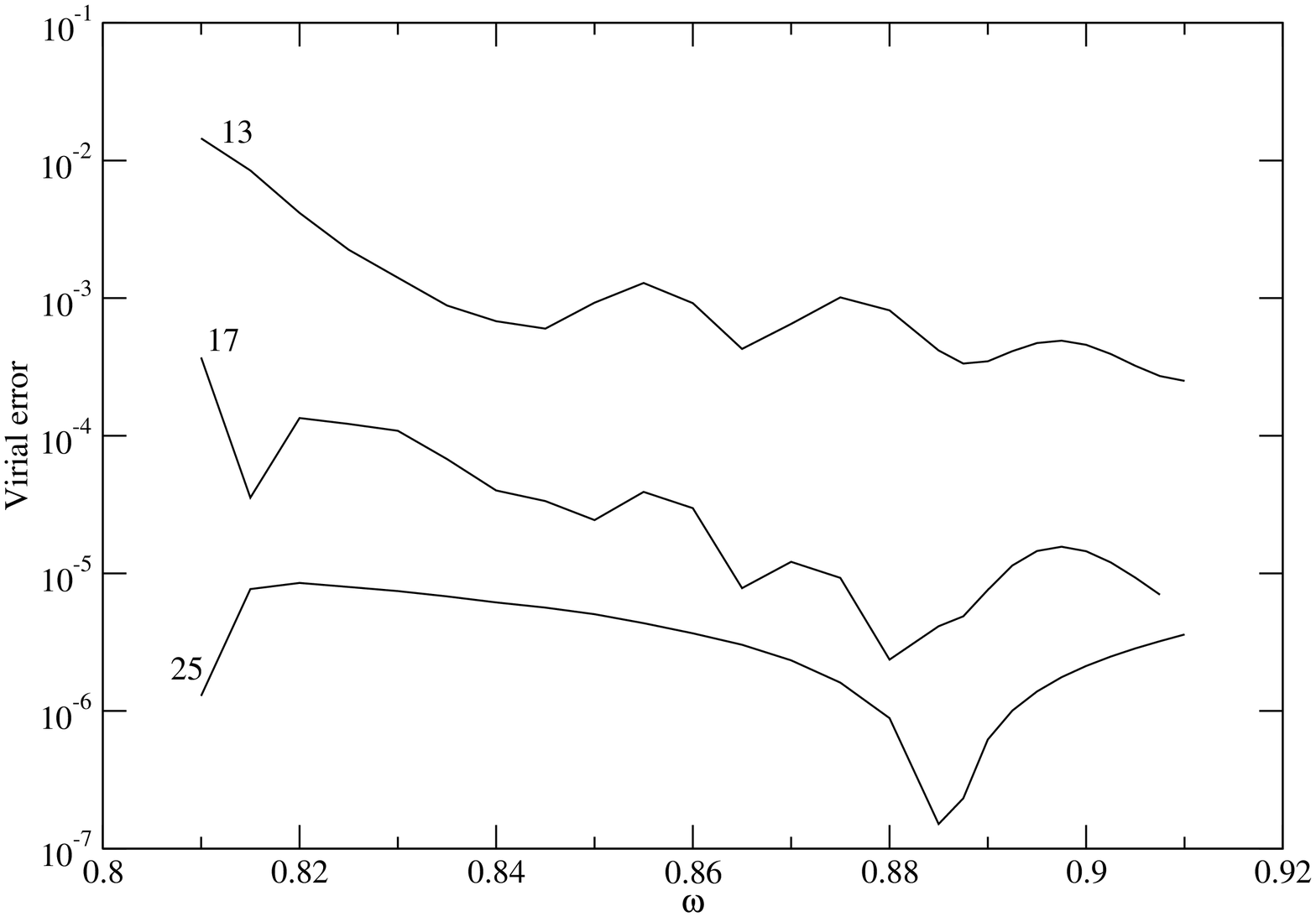} 
\includegraphics[width=0.55\textwidth]{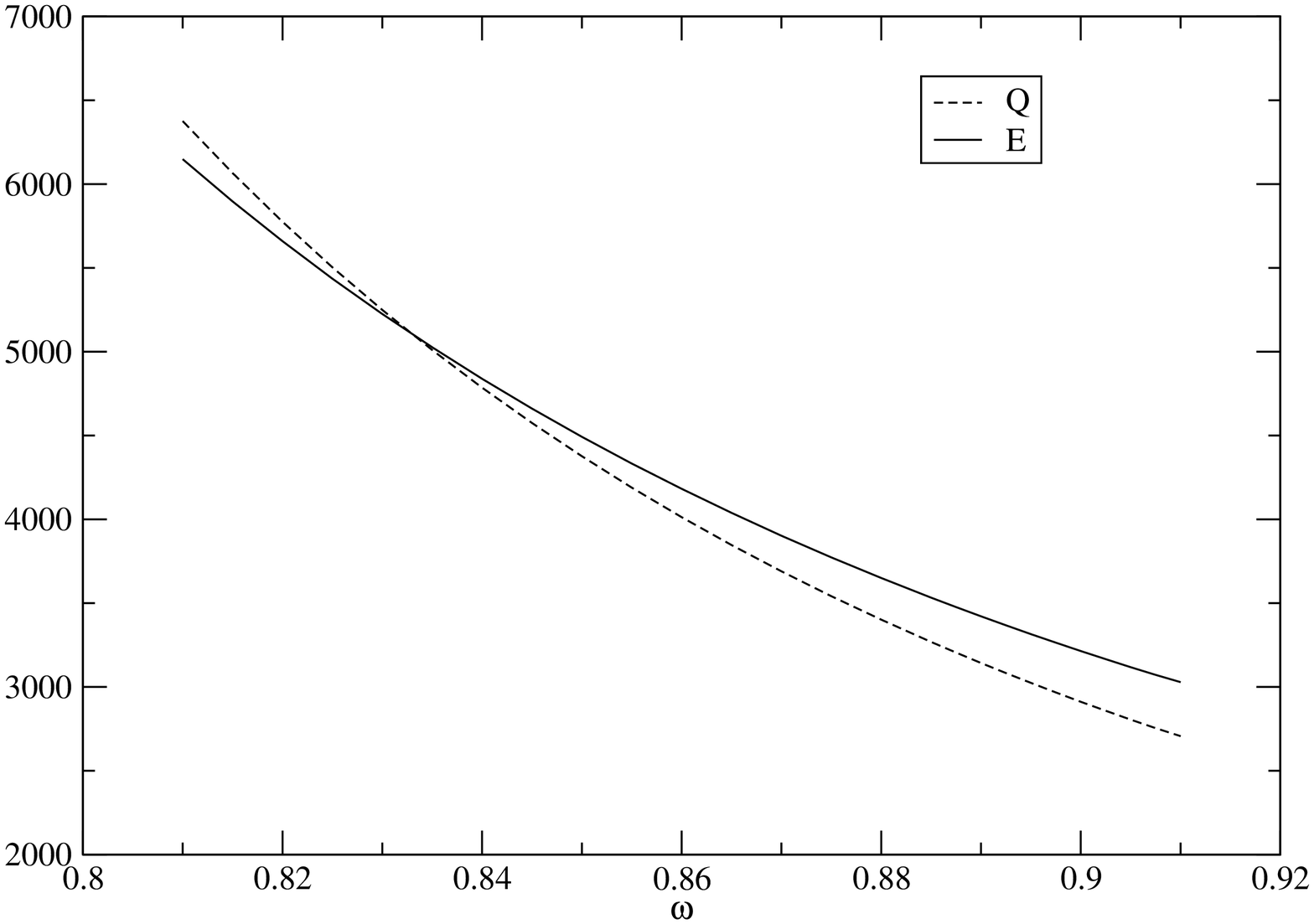} 
\caption{ \label{f:virial_vorton}
Sequence of vortons with different $\omega$. The other parameters are $m=2$, $\lambda_\phi=41.12$, $\lambda_\sigma=40$, $\eta_\sigma=1$ and $\gamma=22.3$. The left panel shows the virial error for three different resolutions. The curves are labeled by the number of points in each dimension. The right panel shows the total energy $E$ and Noether charge $Q$ along the same sequence.}
\end{figure}

The results presented in this section aim at reproducing some of the results obtained in \cite{RaduV08} in the case $m=2$. The other parameters are as follows 
$\lambda_\phi=41.12$, $\lambda_\sigma=40$, $\eta_\sigma=1$ and $\gamma=22.3$. One configuration computed in \cite{RaduV08} is used as an initial guess and a sequence of configurations rotating at different speeds is then constructed by slowly varying $\omega$. The precision reached by the code is asserted by computing the deviation of the solutions with respect to some virial theorem. The virial error is defined as $1- \displaystyle\frac{3\left(\omega^2 {\mathcal N} - E_0\right)}{E_2}$ (see Eq. (6.159) of
\cite{RaduV08}). ${\mathcal N}$, $E_0$ and $E_2$ are integrals of the fields which explicit expressions are given 
by Eqs. (6.154) of \cite{RaduV08}. The virial error is shown in the left panel of Fig. \ref{f:virial_vorton}, as a function of $\omega$, for three different resolutions. The convergence of the virial error is rapid, as expected with spectral method. The error is below $10^{-5}$ in the whole range of $\omega$, for the highest resolution.

\begin{figure}[htb]
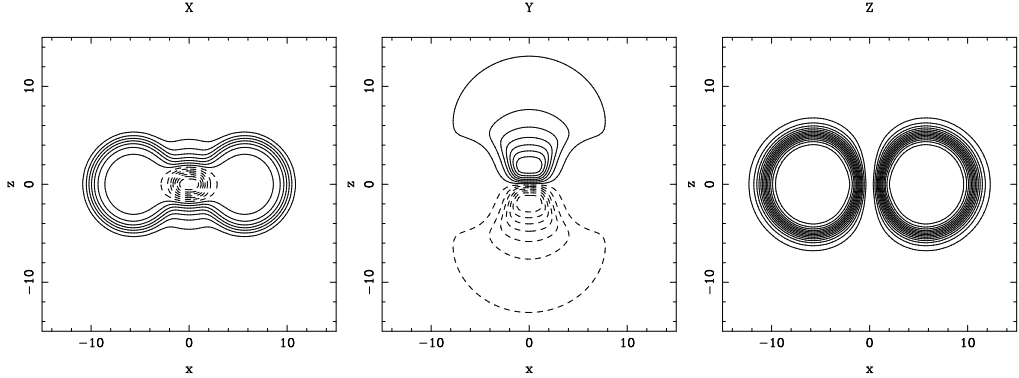

\includegraphics[width=0.32\textwidth]{X.eps} 
\includegraphics[width=0.32\textwidth]{Y.eps} 
\includegraphics[width=0.32\textwidth]{Z.eps}
\caption{ \label{f:fields_vorton}
Fields $X$, $Y$ and $Z$ (from left to right) for $\omega=0.85$. The other parameters are the same as in Fig. \ref{f:virial_vorton}. The continuous (resp. dashed) lines show constant positive (resp. negative) value of the fields.
}
\end{figure}

The right panel of Fig. \ref{f:virial_vorton} shows the total energy $E$ an the Noether charge $Q$ of the vortons, as a function of $\omega$, as defined by Eqs. (6.155) and (6.157) of 
\cite{RaduV08}. The results are consistent with Fig. 30 of \cite{RaduV08}. For the low values of $\omega$, one is limited by the fact that the vortons are bigger and bigger. One would need to use more domains to get to smaller $\omega$. At the higher end, there seem to be a transition from the vorton solutions to solutions for which $Y=0$ in the whole space. This transition prevents the code from reaching as high values of $\omega$ as in \cite{RaduV08}, for a yet unknown reason. Figure \ref{f:fields_vorton} shows contours of constant values for the fields for $\omega=0.85$. Let us finally mention that vortons with $m=1$ were also successfully computed. From the numerical point of view, they mainly differ from the $m=2$ case from the fact that the basis of decomposition for $Z$ is different.

\subsection{Critical collapse}\label{ss:critic}

Critical collapse was first observed in \cite{Chopt93} when computing the evolution of a spherically symmetric massless scalar field in general relativity. In the weak field regime the field disperses at infinity whereas, in the strong field regime, a black hole is formed via gravitational collapse. Just at the threshold between those two cases, another type of solution appears. It is a naked singularity and is called the critical solution. It has many interesting properties and we refer the reader to \cite{GundlMG07} for a review on this subject. The structure of the critical solution for a scalar field collapse has been extensively studied, for instance in \cite{Gundl97, MartiG03}. This section is devoted to the computation of the critical solution by \K.

As stated in \cite{MartiG03}, the critical spacetime can be found as the solution of a 2-dimensional non-linear problem. Coordinates can be chosen such that one spatial variable $x$ goes from $0$ to $1$ and one periodic variable $\tau$ goes from $0$ to $2 \pi$. $\tau$ is the time coordinate. From the geometrical point of view, the space of interest is a cylinder. The relevant class in \K is called {\tt Space\_critic} and is described in detail in Sec. \ref{ss:cylindrical}. In order to take into account some symmetries at $x=0$, this space is separated into two domains.

The solution involves four fields, two describing the matter $U$ and $V$ and, two for the gravitational field, $f$ and $a$. The fields have some symmetries with respect to 
$\tau=\pi$ so that one can restrict $\tau$ to $\left[0, \pi\right[$. More precisely $a\left(\tau+\pi\right)=a\left(\tau\right)$ (idem for $f$) and $U\left(\tau+\pi\right) = - U\left(\tau\right)$ (idem for $V$). Some symmetries are also present at $x=0$ where the metric fields $f$ and $a$ are even function. Some combinations of the matter field are also even and finite at $x=0$ : $\Pi=\left(V+U\right)/\left(2x\right)$ and $\Psi= \left(V-U\right)/\left(2x^2\right)$.

In order to take the symmetry at $x=0$ into account, different sets of unknowns are used in different regions of space. In the inner domain, one works with $\Pi$ and $\Psi$ and in the outer one with $U$ and $V$. Doing so, the various fields are expanded into the appropriate spectral basis of decomposition, as described in Sec. \ref{sss:cylindrical_base}.

In the outer domain, the equations are given by
\bea
\label{e:fouter} x f_{,x} &=& \left(a^2-1\right) f \\
\label{e:aouter} x \left(a^{-2}\right)_{,x} &=& 1 - \left(1+U^2+V^2\right) a^{-2} \\
\label{e:uouter} x\left(f+x\right) U_{,x} &=& f\left[\left(1-a^2\right)U+V\right] - \frac{2\pi}{\Delta} x U_{,\tau} \\
\label{e:vouter} x\left(f-x\right) V_{,x} &=& f\left[\left(1-a^2\right)V+U\right] + \frac{2\pi}{\Delta} x V_{,\tau}
\eea
where $\Delta$ is the period of the solution that will be discussed later.

In the inner domain, one gets 
\bea
\label{e:finner} x f_{,x} &=& \left(a^2-1\right) f \\
\label{e:ainner} x \left(a^{-2}\right)_{,x} &=& 1 - \left(1+2x^2\Pi^2+2x^4\Psi^2\right) a^{-2} \\
\label{e:piinner} x\left(f^2-x^2\right)\Pi_{,x} &=& -\Pi\left(f^2-x^2\right) + f\left(1-a^2\right)\left(f\Pi+x^2\Psi\right) \\
\nonumber & & + f\left(f\Pi-x^2\Psi\right) + \frac{2\pi}{\Delta} x^2 \left(f\Psi_{,\tau} + \Pi_{,\tau}\right) \\
\label{e:psiinner} x\left(f^2-x^2\right)\Psi_{,x} &=& -2\Psi\left(f^2-x^2\right) + f\left(1-a^2\right) \left(f\Psi+\Pi\right) \\
\nonumber & & + f\left(\Pi-f\Psi\right) + \frac{2\pi}{\Delta} \left(f\Pi_{,\tau} + x^2 \Psi_{,\tau}\right).
\eea

The equations are first order differential equations. The variable $\tau$ being periodic, no boundary conditions is needed. The situation with respect to $x$ is more complicated because some of the factors in front of the derivative vanish at the boundaries. The appropriate number of boundary conditions must be determined by a precise examination of the equations.

Let us first turn to the variable $a$. On the side $x=1$ of the cylinder, Eq. (\ref{e:aouter}) is not degenerated, so that it is treated in the standard way, that is standard first order $\tau$-method (see Sec. \ref{ss:weighted}). At the inner boundary however, Eq. (\ref{e:ainner}) is degenerated and becomes $a^2=1$. The equation has to be treated by a $\tau$-method of zeroth order (all the residual equations are kept). In a sense the equation is its own boundary condition.

For the other metric field $f$, in the outer region, the situation is also the standard one. In the inner part, 
Eq. (\ref{e:finner}) is also degenerated and becomes $f\left(a^2-1\right)=0$. As previously seen, this condition is already accounted for by the equation for $a$ at $x=0$. This implies that the degeneracy is not effective and that Eq. (\ref{e:finner}) must be treated by a standard first order 
$\tau$-method. So the field $f$ does require a boundary condition on one side of the cylinder. It is physically motivated and simply is $f\left(x=1\right)=1$.

Given this value of $f$ at the outer side, one of the equations for the matter fields equation (\ref{e:psiinner}) is actually degenerated and equivalent to the regularity condition (20) of \cite{MartiG03} (the sign in \cite{MartiG03} is faulted by a typo). On the inner side equation (\ref{e:piinner}) is degenerated but equivalent to $f^2 \Pi \left(1-a^2\right)=0$ and is already accounted for by the equation for $a$. The other matter equation (\ref{e:psiinner}) is also degenerated and gives a true condition $f\Pi - 3 f^2 \Psi + {2\pi}/\Delta f \Pi_{,\tau}=0$. So, concerning the matter fields, no additional boundary conditions are needed for there are two degeneracy conditions (one on each side), for two fields.

As one is dealing with first order equations, the fields themselves must be matched at the interface between the two domains. This study is slightly technical but it ensures that the constructed global system of equations is well-posed and invertible. The final situation is summarized in Tab. 
\ref{t:critic_situation}. Let us mention that such complicated setting (non-trivial matchings, different order in the $\tau$-method...) can be easily encoded in \K.

\begin{table}
\caption{Construction of a well-posed problem. The columns $x=0$ and $x=1$ summarize the behavior of the equations at the boundaries. The columns labeled inner and outer give the order of the associated $\tau$-method. The matching conditions at the interface are also given. More details about this can be found in the core of the text.}
\begin{tabular}{c||c|c|c|c|c}\label{t:critic_situation}
Field & $x=0$ & inner& matching & outer& $x=1$\\
\hline
$a$ & dege. $\Rightarrow a^2=1$ & $0^{\rm th.}$ & $a^{-}=a^{+}$ & $1^{\rm st}$ & non dege.  \\
\hline
$f$ & dege. $\Leftrightarrow a^2=1$ & $1^{\rm st}$& $f^{-}=f^{+}$ & $1^{\rm st}$ & $f=1$  \\
\hline
matter &$\Pi$ : dege. $\Leftrightarrow a^2=1$ & For $\Pi$ : $1^{\rm st}$ & $2 x \Pi^{-} = V^{+} + U^{+}$
&$U$ :  $1^{\rm st}$ & $U$ : non dege. \\
      & $\Psi$ : dege. & $\Psi$ : $0^{\rm th}$ & $2 x^2 \Psi^{-} = V^{+} - U^{+}$ 
& $V$ :  $0^{\rm st}$ & $V$ : dege. 
\end{tabular}
\end{table}

Along with the field equations, there is also a global condition that constraints the value of $\Delta$ and that demands that the mode corresponding to $\cos\left(2\tau\right)$ in the expansion of $f$ vanishes at $x=0$. This last condition falls into the category of the integral equations discussed in 
Sec. \ref{ss:integral_eqs}. It is believed that the equations are very sensitive to the initial guess and that they will fail to converge to the critical one if one starts too far from it. This is the reason for using the solution computed by \cite{MartiG03} as a starting point of the Newton-Raphson iteration. The errors that appear when passing data from one code to the other introduce enough numerical noise to make the loop do a few iterations.

\begin{figure}[htb]
\includegraphics[width=0.6\textwidth]{error_delta.eps} 
\includegraphics[width=0.6\textwidth]{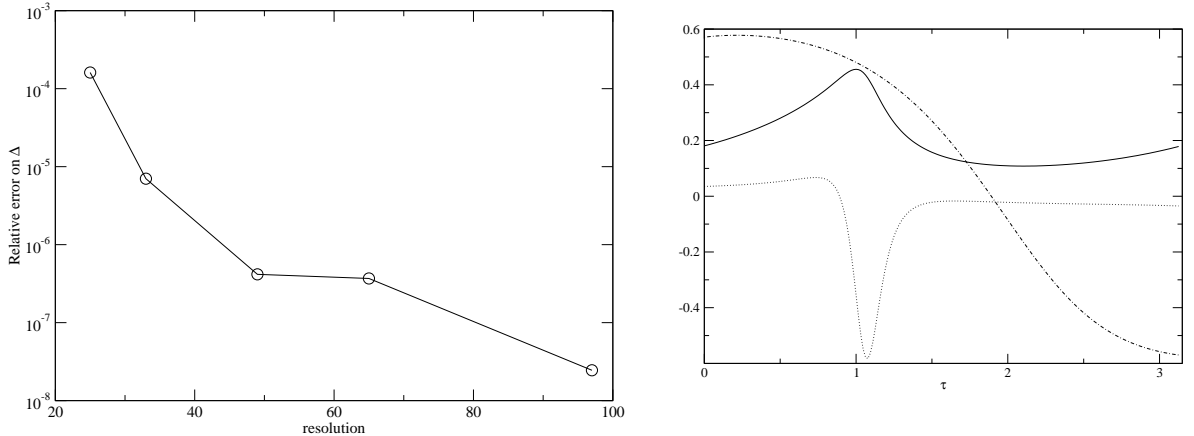} 
\caption{ \label{f:critic}
The left panel shows the relative error in computing the period $\Delta$. The exact value is assumed to be given by Eq. (58) of \cite{MartiG03}. The convergence is shown with respect to the number of points in each dimension. The right panel shows the values of $f\left(x=0\right)$ (continuous line), 
$\Psi \left(x=0\right)/300$ (dotted line) and $U\left(x=1\right)$ (dash-dot line), as a function of $\tau$.}
\end{figure}

The precision of the code is assessed by comparing the value of the period $\Delta$ to the one given by Eq. (58) of \cite{MartiG03}. The relative difference between the two is shown in the left panel of Fig. \ref{f:critic}, as a function of the number of points in each dimension. The convergence is satisfactory even though it is difficult to state that the evanescent regime has been reached. Let us note that the convergence of the solution is significantly slower than for other problems, an accuracy of $10^{-8}$ being reached with as many as 96 points in both dimensions. This is expected, given the strong gradients appearing in the fields. Those gradients are responsible for a rather low decay of the coefficients of the spectral expansions. This was already observed in \cite{MartiG03} and can be seen in their Fig. 8 and 9. As an illustration, some fields at both sides of the cylinder are shown in the right panel of Fig. \ref{f:critic}. This plot is the equivalent of Fig. 4 of \cite{MartiG03}, except for the definition of the coordinate $\tau$. In \cite{MartiG03}, the temporal coordinate $\tau$ goes from 0 to $\Delta$, whereas in this work, $\tau/\Delta$ is used so that our coordinate goes from $0$ to $\pi$, once the symmetry at half-period is taken into account.

\subsection{Binary black holes}\label{ss:hkv}

There is a long history of works that aim at computing the structure of spacetimes that contain a system of binary black holes. Amongst the many reasons to study such systems, one will mention the fact that they are known to be good emitters of gravitational waves and so are one of the main target for the gravitational wave detectors currently in operation \cite{SathyS09}. Binary systems have also a great interest in the context of galaxy formation. Indeed, it is believed that nowadays galaxies have been formed by successful mergers of smaller galaxies. When the merger occurs, the black holes present at the center of the smaller galaxies will become bounded and form a binary system \cite{MerriM05}.

Under the influence of gravitational radiation, two black holes will not remain on closed orbits but rather spiral towards each other, until they merge into a single object. This process occurs in the strong field regime of gravitation and must be described in the context of general relativity. Most of the simulations are performed in the 3+1 formalism where a splitting of space and time is introduced \cite{York79}. The main effect of this splitting is to separate Einstein's equations into two sets: (i) the constraint equations that do not involve time (ii) and the evolution equations that contains Dalembert type operators. Doing so, the simulation of a binary system proceeds in two separate steps. First, one needs to produce a initial configuration that must satisfy the constraint equations and that represents the physics of interest as accurately as possible. This is known as the initial data problem \cite{Cook05}. In a second step the behavior of the fields at latter times is obtained by using the evolution equations. Let us mention that if the constraint equations are fulfilled at the initial time, they are verified at all time steps, if the evolution equations are used properly.

In this section, one is interested in preparing an initial configuration that represents two black holes in quasi-circular orbit along the same lines as in \cite{CaudiCGP06}. Even if true circular orbits cannot exist due to gravitational radiation, it is believed to be a rather good approximation, at least when the black holes are relatively far apart. Technically this is done by imposing the existence of an helical Killing vector (see for instance \cite{BonazGM97, GourgGBM02} and references therein). Another approximation usually done in this context is the so-called conformal flatness approximation (see below for a precise definition). This has more to do with the substantial mathematical simplification that results than with any physical motivation. It was once believed that the conformal flatness approximation would minimize the amount of spurious radiation in the data sets but it turned out not to be the case. Nevertheless, the approximation leads to a consistent mathematical problem, especially as far as the behaviors of the fields at infinity are concerned \cite{GourgGBM02}.

In the 3+1 formalism, the 4-dimensional metric $g_{\mu \nu}$ that describes the system is expressed in terms of purely spatial quantities. More precisely the line element is given by
\be
{\rm d}s^2 = g_{\mu \nu} {\rm d}x^\mu {\rm d}x^\nu = - \left(N^2 - N^i N_i\right) {\rm d}t^2 +
2 N_i {\rm d}t {\rm d}x^i + \gamma_{ij} {\rm d}x^i {\rm d}x^j
\ee
where the Greek indices are 4-dimensional and run from $0$ to $3$ and the Latin ones are purely spatial and go from $1$ to $3$. The 3+1 quantities that describe the geometry are one scalar $N$ the lapse, one vector $N^i$ the shift (three independent functions) and one spatial metric $\gamma_{ij}$ (six independent functions). The conformal flatness approximation states that the spatial metric relates to the flat metric by a single scalar conformal factor $\Psi$. More precisely one demands that
\be
\gamma_{ij} = \Psi^4 f_{ij},
\ee
where $ f_{ij}$ is the flat metric. This condition is only an approximation and it is not believed to be exactly true.

In this setting, the unknowns are the fields $N$, $\Psi$ and $N^i$. The associated equations are obtained from Einstein equations by enforcing stationarity (i.e. the time derivative are set to zero). It gives a set of five elliptic equations:
\bea
\label{e:traceevol}
D_i D^i N &=& - 2 \frac{D_i \Psi D^i N} {\Psi} + N \Psi^4 A_{ij} A^{ij} \\
\label{e:hamilton}
D_i D^i \Psi &=& - \frac{\Psi^5}{8} A_{ij} A^{ij} \\
\label{e:momentum}
D_j D^j N^i + \frac{1}{3} D^i D_j N^j &=& 2 A^{ij} \left(D_j N -6 N \frac{D_j \Psi}{\Psi}\right)
\eea
where $D$ is the covariant derivative with respect to the flat metric. $f_{ij}$ is also used to raise and lower indices of tensors. $A^{ij}$ is the conformal extrinsic curvature tensor and represents the way the three metric is embedded in the 4-dimensional geometry. From the mathematical point of view it is defined as being
\be
\label{e:defa}
A^{ij} = \frac{1}{2N} \left(D^i N^j + D^j N^i - \frac{2}{3} D_k N^k f^{ij}\right).
\ee

The elliptic equations must be supplemented by appropriate boundary conditions. At infinity, it is demanded that flat spacetime is recovered which implies that
\be
\label{e:infinity}
N=1 \quad , \quad N^i=0 \quad {\rm and} \quad \Psi=1 \quad {\rm when} \quad r\rightarrow \infty.
\ee
The inner boundary conditions are enforced on two spheres that represent the black holes themselves. They are basically obtained by demanding that those two spheres are apparent horizons in equilibrium (see \cite{GourgJ05} for a review). Those inner boundary conditions enforce the presence and the physical properties of the holes. On the sphere ${\mathcal S}_a$ ($a=1$ or $2$), one gets
\bea
\label{e:bclapse} \left. N \right|_{\mathcal S_a}  &=& n_0 \\
\label{e:bcpsi} \left. \frac{\partial \Psi}{\partial_r} + \frac{\Psi}{2 r} \right|_{\mathcal S_a} &=& - \frac{\Psi^3}{4} A_{ij} s_a^i s_a^j\\
\label{e:bcshift} \left. N^i \right|_{\mathcal S_a} &=& \frac{n_0}{\Psi^2} s_a^i + \Omega M_a^i + \Omega_a m_a^i.
\eea

Equation (\ref{e:bclapse}) is just a choice of time coordinate on the spheres and $n_0$ is a constant (in this section one takes $n_0=0.1$). Equation (\ref{e:bcpsi}) translates the fact that the spheres are apparent horizons. $s_a^i$ is the unit vector normal to the sphere which reads $\left(x_a/r_a, y_a/r_a, z_a/r_a\right)$ in Cartesian coordinates centered on the sphere $a$. Equation (\ref{e:bcshift}) states that the spheres are in equilibrium and also contains information about the state of rotation of the holes. $\Omega$ is the orbital velocity of the system and $M_a^i$ the constant vector $\left(0, X_a, 0\right)$ where $X_a$ is the coordinate distance between the center of mass and the center of the hole $a$. This part of the boundary condition accounts for the orbital motion of the holes. $\Omega_a$ is the local rotation rate of the black hole $a$ and it is associated to the local vector $m_a^i = \left(-y_a, x_a, 0\right)$. This last part accounts for the spins of the objects.

$\Omega$ and the two $\Omega_a$ are global unknowns that must be constrained by additional equations, as discussed in Sec. \ref{ss:integral_eqs}. The proper value of the orbital velocity is obtained by demanding that two integral quantities, the Komar mass and the ADM mass are equal. This is closely linked to a virial theorem, as discussed in \cite{GourgGBM02}.The two masses are defined as surface integrals at infinity
\bea
\label{e:komar}
M_{\rm Komar} &=& \frac{1}{4\pi} \oint_\infty D^i N {\rm d}S^i\\
\label{e:adm}
M_{\rm ADM} &=& \frac{-1}{2\pi} \oint_\infty D^i \Psi {\rm d}S^i.
\eea
In this work, the local rotation rates are determined by demanding that the black holes are not rotating and so that their spin $S_a$ vanish. The individual spins can be defined by integrals on the spheres themselves
\be
S_a = \frac{1}{8\pi} \oint_{\mathcal S} \Psi^6 A_{ij} m_a^i {\rm d}S^j.
\ee
The total angular momentum is defined by a similar integral, taken at infinity, and reads
\be
J = \frac{1}{8\pi} \oint_\infty A_{ij} m^i {\rm d}S^j,
\ee
where $m^i =  \left(-Y, X, 0\right)$. $X$ and $Y$ are the Cartesian coordinates with respect to the center of mass.

This problem has been solved in the context of \K by using the bispherical space implemented by the class
{\tt Space\_bispheric} (see Sec. \ref{ss:bispheric}). The input parameters are the radii of the spheres (taken equal in this particular case) and their separation. A Cartesian tensorial basis of decomposition is used. The unknown are the fields $N$, $\Psi$ and $N^i = \left(N^x, N^y, N^z\right)$ and the global quantities $\Omega$ and $\Omega_a$. $A_{ij}$ is defined as being a definition in terms of the unknowns (given by Eq. (\ref{e:defa})). The various vectors appearing in the equations, like the $m_i$ are passed to \K as constants. The equations are given by Eqs. (\ref{e:traceevol}, \ref{e:hamilton}, \ref{e:momentum}) and the boundary conditions enforced on both the spheres and at infinity. Matching of the fields and their normal derivatives at the boundaries between the domains is imposed. The three integral equations constraining the global quantities are also passed to the solver.

\begin{figure}[htb]
\includegraphics[width=0.8\textwidth]{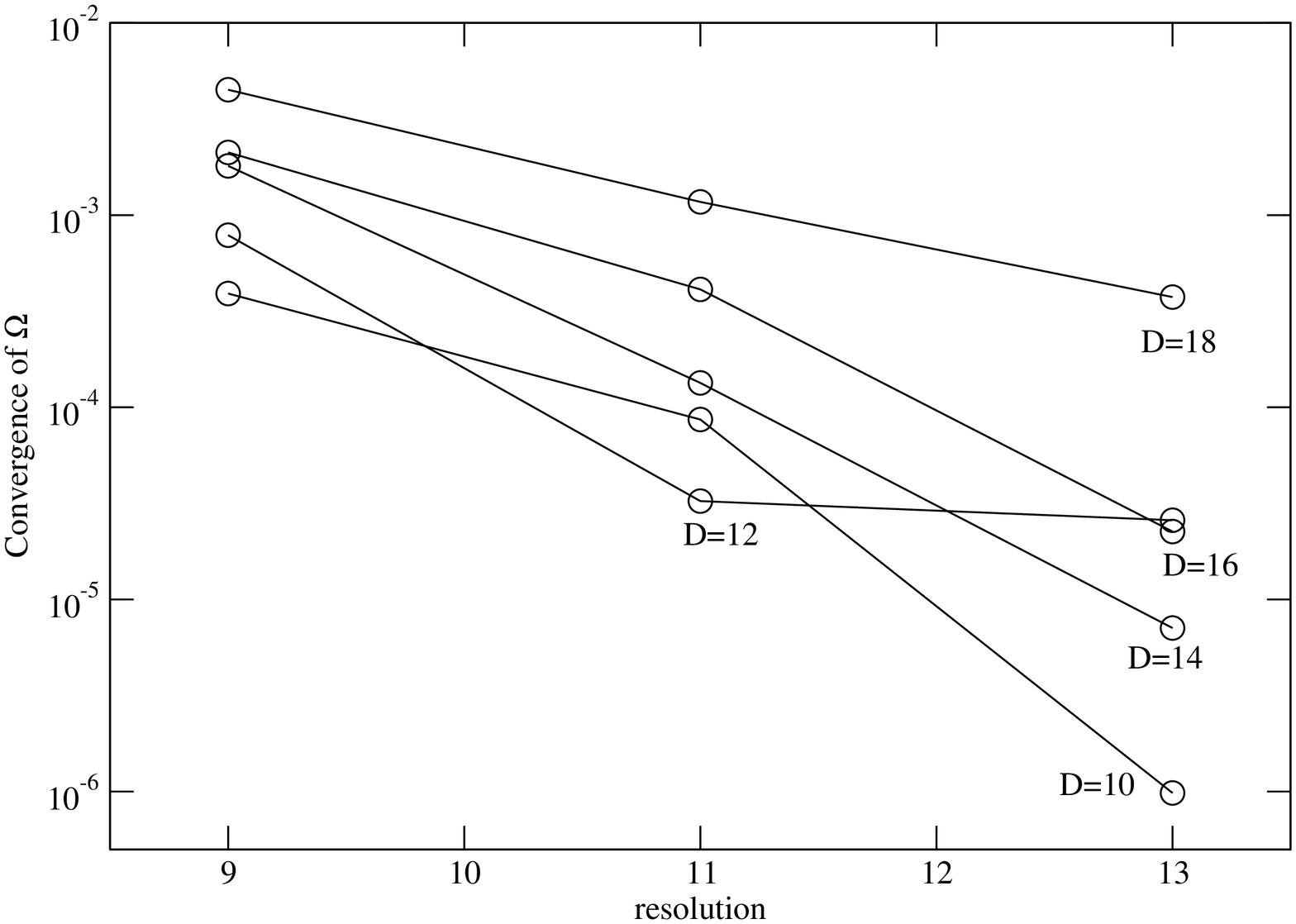} 
\caption{ \label{f:conv_hkv}
Convergence of $\Omega$ as a function of resolution (compared to the best available value obtained with 15 points in each dimensions). The convergence is somewhat faster for small separations.
}
\end{figure}

Contrary to the other cases presented in this paper, the binary black hole configurations are really three-dimensional. This implies that the computational task is somewhat harder. Configurations with four different resolutions have been computed with 9, 11, 13 and 15 points, respectively, in all dimensions and in all the domains. Let us mention that the size of the Jacobian for the higher resolution is just less than 100,000. Configurations for five different separations are computed. It appears that the values of the fields depend strongly from the value of $\Omega$ which must be obtained with a good accuracy. Such accuracy can be measured by looking at the convergence of $\Omega$ as the function of the number of points as in Fig. \ref{f:conv_hkv}. The convergence is shown by taking the difference between the value of $\Omega$ found for given resolution and the one for the best available resolution ($15$ points in each dimension in this case).

\begin{figure}[htb]
\includegraphics[width=0.6\textwidth]{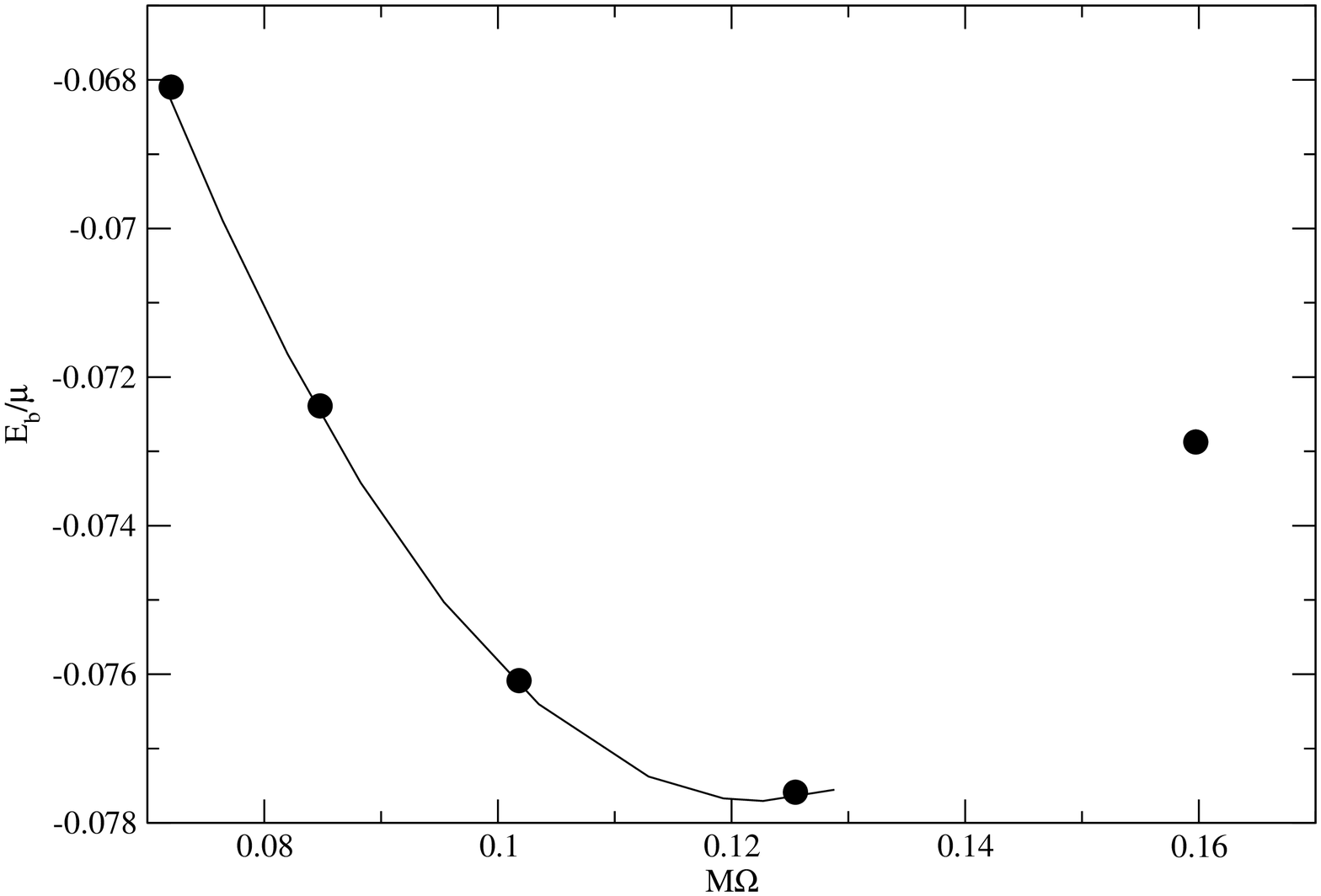} 
\includegraphics[width=0.6\textwidth]{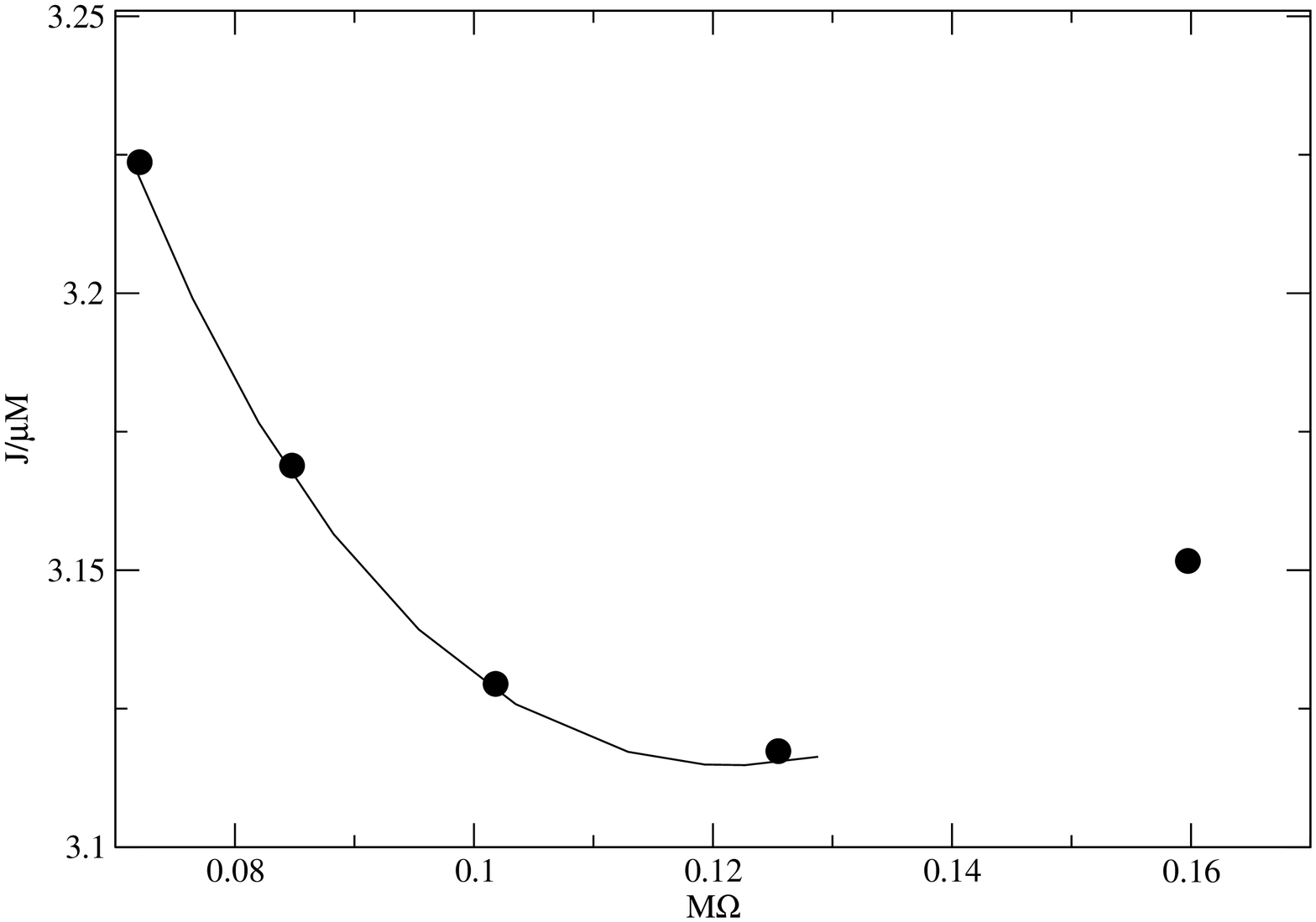}
\caption{ \label{f:curves_hkv}
Binding energy (left panel) and total angular momentum (right panel), as a function of $M\Omega$. The circles denote the values obtained by \K and the solid curves are data taken from \cite{CaudiCGP06}.
}
\end{figure}

The physical quantities are usually given in their adimensional form. This can be done by making use of the area mass of the black holes: $m=\sqrt{A/16\pi}$. $A$ is the area of the hole which is given by an integral on the horizon (i.e. the sphere) :
\be
A = \int_{\mathcal S} \Psi^4 {\rm d}S.
\ee
The area mass of the system is simply $M=m_1 + m_2$ and is constant along a sequence of varying separation (see for instance \cite{GrandGBM02}). It follows that $M$ is the total mass of the system when the black holes are infinitely separated. One can then define the binding energy of the system as $E_b = M_{\rm ADM} - M$. The reduced mass is defined in the usual manner and in the case of equal mass black holes is given by $\mu=M/4$. The adimensional binding energy $E_b/\mu$ and the adimensional total angular momentum $J/M\mu$ are shown on Fig. \ref{f:curves_hkv}, as a function of the adimensional orbital velocity $M\Omega$. The values obtained by \K are compared to the data from \cite{CaudiCGP06}. The agreement is very good, especially considering that the variations of $J$ (resp. $E_b$) are small compared to the values of $J$ (resp $M_{\rm ADM}$) itself.

\begin{figure}[htb]
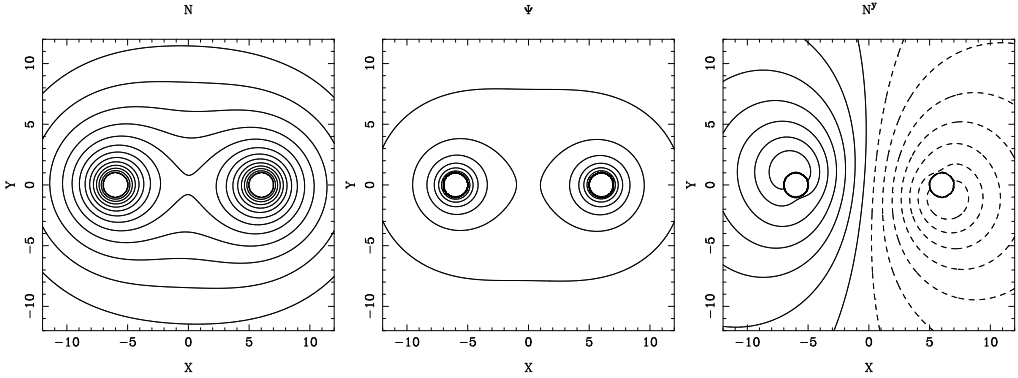

\includegraphics[width=0.32\textwidth]{lapse_hkv.eps} 
\includegraphics[width=0.32\textwidth]{conf_hkv.eps} 
\includegraphics[width=0.32\textwidth]{shift_hkv.eps}
\caption{ \label{f:fields_hkv}
Values of the lapse function (left panel), the conformal factor (center panel) and the component $N^y$ of the shift (right panel), in the orbital plane $z=0$. The coordinate separation is 12. The continuous (resp. dashed) lines show constant positive (resp. negative) value of the fields.
}
\end{figure}

Finally, contours of constant values of some fields are shown in Fig. \ref{f:fields_hkv}, in the orbital plane (the plane $z=0$) for a separation of 12. The locations of the spheres are clearly visible.

\subsection{Kerr problem}\label{ss:kerr}

In this section, one aims at recovering the exact solution for a single stationary rotating black hole, along the lines of \cite{VasseNJ09}. It is called the Kerr solution and is known to be analytic, at least for some choices of coordinates \cite{Kerr63}. This is however not the case for the formalism that is used here. The starting point is the same as the one used in the binary system case in Sec. \ref{ss:hkv}. An orthonormal spherical tensorial basis of decomposition is used. Doing so and for this problem, all the fields will independent of $\varphi$, thus recovering the fact that the solution is axisymmetric.

In this section, and as opposed to what is done in Sec. \ref{ss:hkv}, one aims at recovering an exact solution and so one needs to remove the conformal flatness approximation. Indeed, even for a single black hole, it is known that their exists no choice of coordinates for which the conformal metric is flat \cite{GaratP00}. That being said, the unknowns are the same as in Sec. \ref{ss:hkv} : two scalars the lapse $N$ and the conformal factor $\Psi$ and one vector, the shift $N^i$, to which one must add the conformal spatial metric $\tilde{\gamma}_{ij}$ itself. The metric is a rank-2 symmetric tensor and so has six different components. By definition of the conformal factor, $\tilde{\gamma}_{ij}$ is such that its determinant is one. As will be seen in the following, the conformal metric still contains some gauge degrees of freedom.

The constraint equations along with the trace of the evolution ones can be written as
\bea
\label{e:kerr_traceevol}
\tilde{D}_i \tilde{D}^i N &=& - 2 \frac{\tilde{D}_i \Psi \tilde{D}^i N} {\Psi} + N \Psi^4 A_{ij} A^{ij} \\
\label{e:kerr_hamilton}
\tilde{R} - 8 \frac{\tilde{D}_i \tilde{D}^i \Psi}{\Psi} &=& \Psi^4 A_{ij} A^{ij} \\
\label{e:kerr_momentum}
\tilde{D}^j A_{ij} &=& - 6 A_{ij} \frac{\tilde{D}^j \Psi}{\Psi} 
\eea
where $A^{ij}$ is the extrinsic curvature tensor which is given by 
\be
\label{e:kerr_defa}
A^{ij} = \frac{1}{2N} \left(\tilde{D}^i N^j + \tilde{D}^j N^i - \frac{2}{3}\tilde{D}_k N^k \tilde{\gamma}^{ij}\right)
\ee
where $\tilde{D}$ denotes the covariant derivative associated to $\tilde{\gamma}_{ij}$ and $\tilde{R}$ is its scalar Ricci tensor. Equations (\ref{e:kerr_traceevol}) to 
(\ref{e:kerr_defa}) are the equivalent of Eqs. (\ref{e:traceevol}) to (\ref{e:defa}), in the non-conformally flat case.

At infinity one demands that flat spacetime is recovered which gives $N=1$, $\Psi=1$ and $N^i=0$, as in Sec. \ref{ss:hkv}. The inner boundary conditions are obtained in the same manner as in the binary case except that they must be written with a general (non-flat) spatial three metric. One then gets, on the sphere 
${\mathcal S}$
\bea
\label{e:kerr_bclapse} \left. N \right|_{\mathcal S}  &=& n_0 \\
\label{e:kerr_bcpsi} \left. 4  \tilde{s}^i \frac{\tilde{D}_i \Psi}{\Psi} + \tilde{D}_i \tilde{s}^i \right|_{\mathcal S} &=& - \Psi^2 A_{ij} \tilde{s}^i \tilde{s}^j\\
\label{e:kerr_bcshift} \left. N^i \right|_{\mathcal S} &=& \frac{n_0}{\Psi^2} \tilde{s}^i + \Omega m^i.
\eea
Equations (\ref{e:kerr_bclapse}) to (\ref{e:kerr_bcshift}) are the equivalent of Eqs. (\ref{e:bclapse}) to (\ref{e:bcshift}), in the non-conformally flat case and for a single black hole. $\tilde{s}$ is the unit outgoing normal to the sphere, with respect to the conformal metric and $\Omega$ the rotation parameter of the hole.

The evolution equations, written in the stationary case, can be seen as equations for the spatial metric and they are
\bea
\label{e:kerr_evol}
\tilde{D}_i \tilde{D}_j N - 2 \frac{\tilde{D}_i N \tilde{D}_j \Psi}{\Psi} - 2 \frac{\tilde{D}_j N \tilde{D}_i \Psi}{\Psi} + 2 \tilde{\gamma}_{ij} 
\frac{\tilde{D}_k N \tilde{D}^k\Psi}{\Psi} - N \tilde{R}_{ij} - 6 N \frac{\tilde{D}_i\Psi  \tilde{D}_j\Psi} {\Psi^2} &&\\
\nonumber
 + 2 N \frac{\tilde{D}_i\tilde{D}_j\Psi}{\Psi} + 2 N \tilde{\gamma}_{ij} \frac{\tilde{D}_k\tilde{D}^k\Psi}{\Psi} + 2 N \tilde{\gamma}_{ij}\frac{\tilde{D}_k\Psi  \tilde{D}^k\Psi} {\Psi^2} + 2 N \Psi^4 A_{ik} A_j^k  &&\\
\nonumber
- \Psi^4 \left(4 A_{ij} N^k \frac{\tilde{D}_k\Psi} {\Psi} + N^k \tilde{D}_k A_{ij} + A_{ik} \tilde{D}_j N^k + A_{jk} \tilde{D}_i N^k\right) = 0&&,
\eea
where $\tilde{R}_{ij}$ is the Ricci curvature of $\tilde{\gamma}_{ij}$. It is the curvature that contains the second order operators in terms of the metric (i.e. the expression of $\tilde{R}_{ij}$ involves terms that look like $\Delta \left(\tilde{\gamma}_{ij}\right)$).

Equations (\ref{e:kerr_evol}) are obviously symmetric and so represent six components. However, if used as such, they do not lead to a well-posed problem because they are not all independent. The construction of a well-behaved problem is rather tricky in this case and is related to both the inner conditions for the spatial metric and the choice of gauge.

The problem of knowing what boundary conditions must be enforced on the metric on the horizon is a very contemporary one and several proposals have been made in the literature. However only one has been successfully applied so far and it is known as the {\em no-boundary} treatment \cite{VasseNJ09}. The basic idea is to assume that
Eqs. (\ref{e:kerr_evol}) are somewhat degenerate on the horizon (like some equations of Sec. \ref{ss:critic}) and so that they require less boundary conditions. If the degeneracy is not strictly demonstrated, there are some good reasons to believe that it is true, at least at the first order 
(see section IV-B of \cite{VasseNJ09}). However the code used in \cite{VasseNJ09} does not use Eqs. (\ref{e:kerr_evol}) as such but a variation of them that aims at imposing that the spatial metric fulfills a particular gauge known as the Dirac gauge. This gauge can be written as
\be
\label{e:dirac}
{\mathcal D}_i \tilde{\gamma}^{ij}=0,
\ee
where ${\mathcal D}$ denotes the covariant derivative with respect to the flat metric. The gauge is used to simplify the explicit expression of $\tilde{R}_{ij}$.

When the full set of Eqs. (\ref{e:kerr_evol}) is used, a full no-boundary treatment leads to a non-invertible system of equations because the gauge has to be imposed at some point. The idea is to use Eq. (\ref{e:dirac}), or at least some part of it, as a boundary condition for the metric. From the numerical experiments conducted with \K, it is found that a way of recovering the Kerr spacetime is to use only the component $\theta$ of Eq. (\ref{e:dirac}) as a boundary condition for 
the component $\left(\theta,r\right)$ of Eqs. (\ref{e:kerr_evol}). The other components of Eqs. (\ref{e:kerr_evol}) are treated as degenerate without any boundary condition. It is somewhat surprising to observe that imposing only one component of the Dirac gauge and only on the horizon is sufficient to obtain the full gauge in whole space, as will be checked a posteriori. Finally one has to note that the conformal metric $\tilde{\gamma}_{ij}$ must be, by definition of $\Psi$, such that its determinant is unity: ${\rm det}\left(\tilde{\gamma}\right)=1$. This is an additional equation on the components of the metric and it is used in place of the $\left(\varphi, \varphi\right)$ component of Eqs. (\ref{e:kerr_evol}).

To summarize the situation, the equations that one uses when solving this problem in the context of \K are, as far as the metric is concerned:
\begin{itemize}
\item the $\left(r,r\right)$,  $\left(r,\varphi\right)$, $\left(\theta,\theta\right)$ and $\left(\theta,\varphi\right)$ components of Eq. (\ref{e:kerr_evol}), without any boundary conditions near the horizon (i.e. a $\tau$-method of first order is used in the inner domain).
\item the $\left(r, \theta\right)$ component of Eq. (\ref{e:kerr_evol}) is used in the standard way, with the component $\theta$ of Eq. (\ref{e:dirac}) as an inner boundary condition.
\item the condition on the determinant of the metric provides the last equation. As it contains no derivative, it is solved by a $\tau$-method of zeroth order.
\end{itemize}
At spatial infinity, one demands that the metric goes to the flat one, which is possible only in the single black hole case (for a binary system, gravitational waves would probably forbid this).

The reason why the setting described above leads to a system of equations that behaves correctly and why it is the only combination that does so, at least in the context of this paper, is currently not known. Further studies and deeper mathematical understanding of the system is probably required but beyond the scope of this work.

\begin{figure}[htb]
\includegraphics[width=0.8\textwidth]{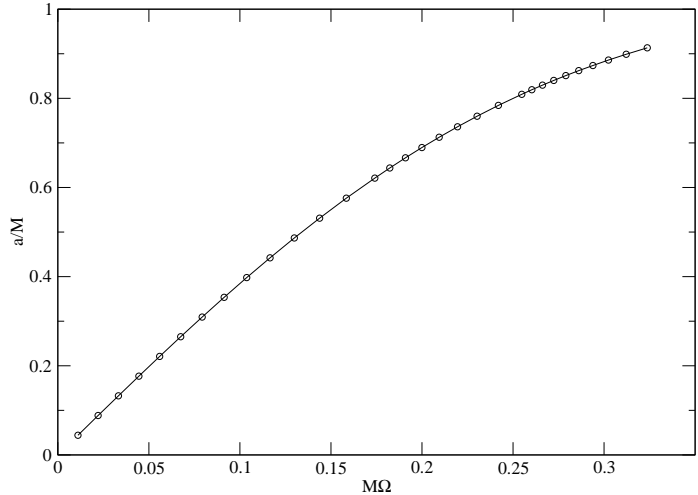} 
\caption{ \label{f:a_f_ome}
Value of the Kerr parameter as a function $M\Omega$ for the high resolution configurations. The values for lower resolutions lies almost exactly on the same curve.
}
\end{figure}

\begin{figure}[htb]
\includegraphics[width=0.6\textwidth]{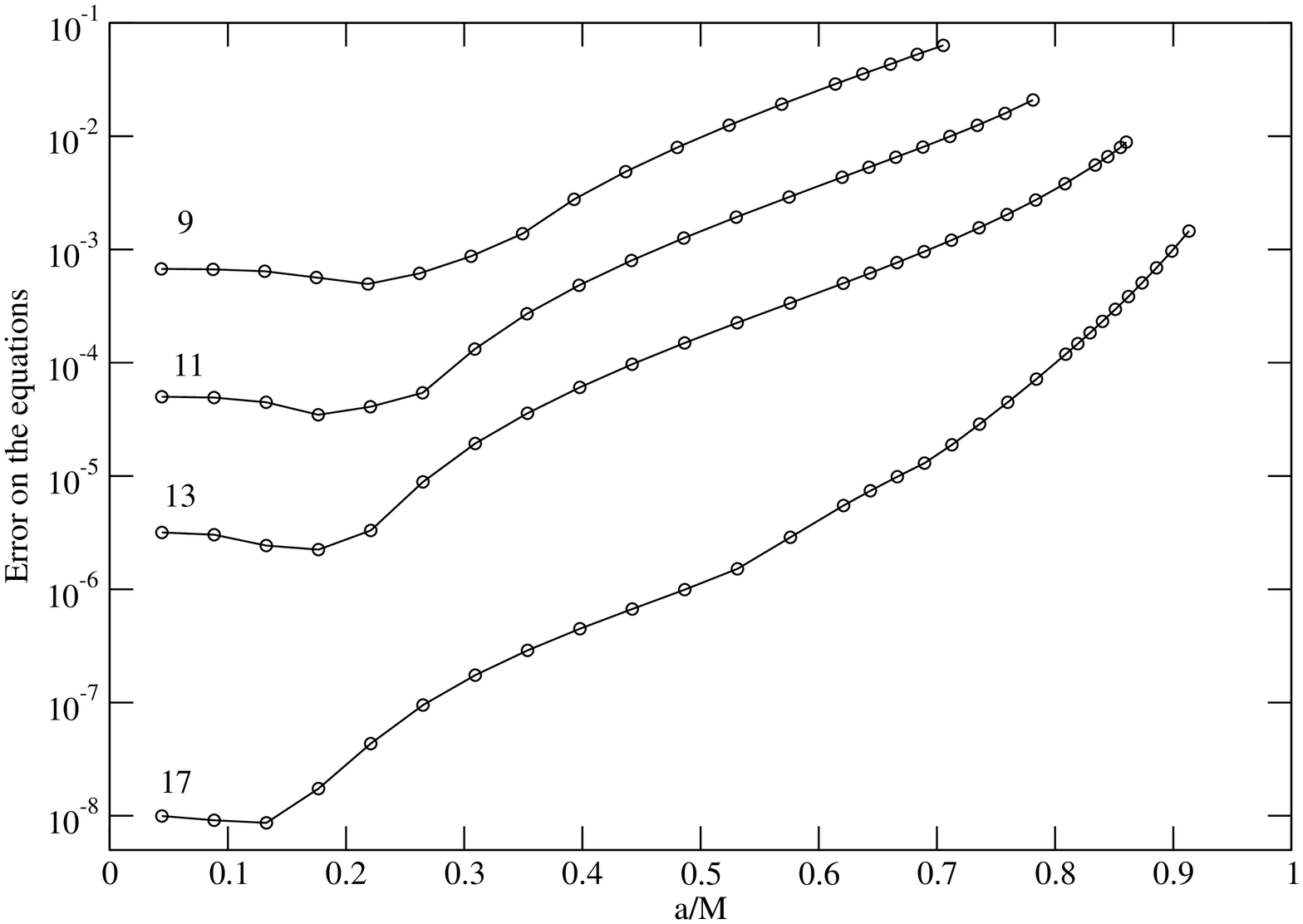} 
\includegraphics[width=0.6\textwidth]{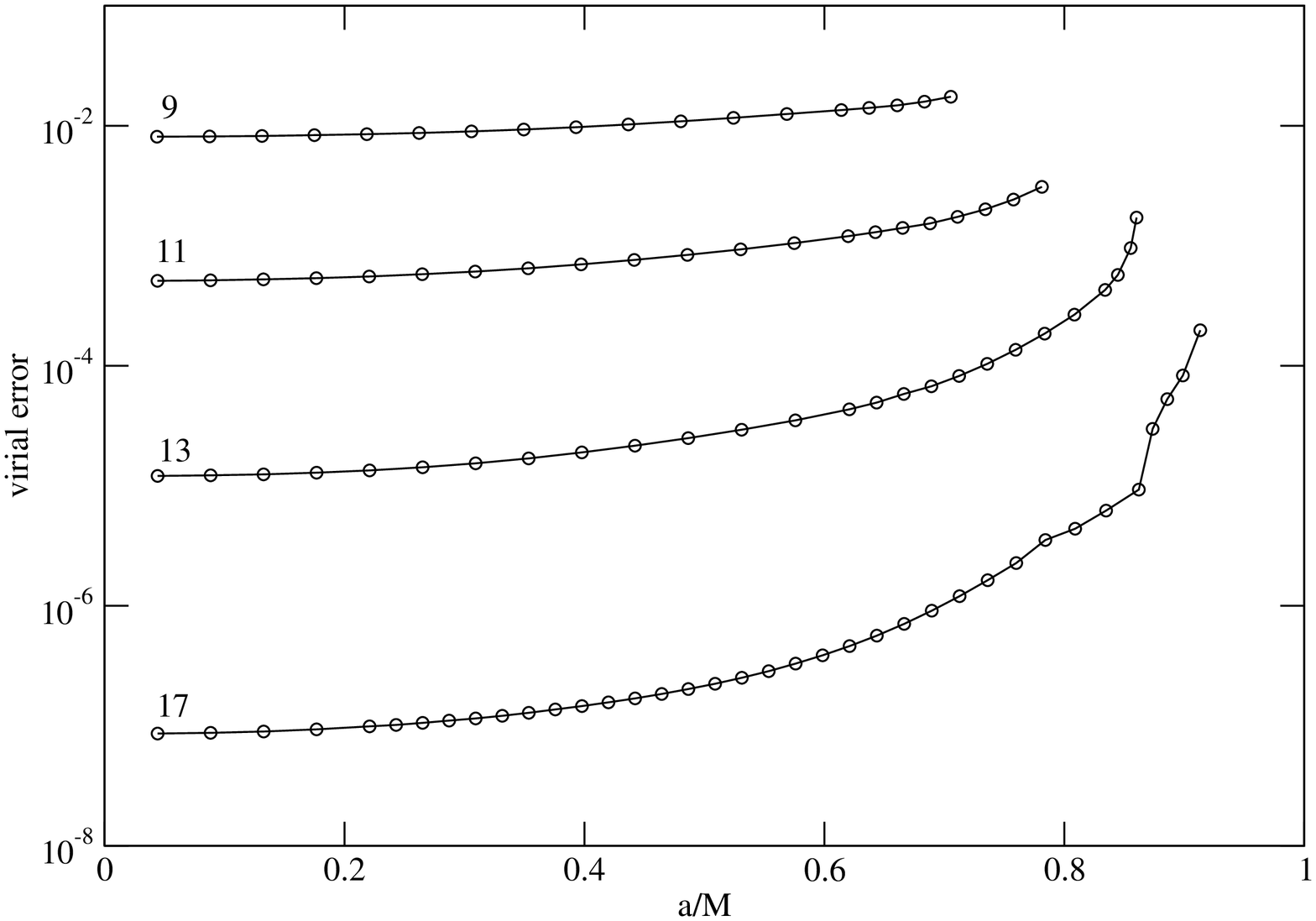}
\caption{ \label{f:kerr_errors}
Precision reached by the code as a function of $a/M$, for the four different resolutions. The left panel shows the maximal error on the full set of equations and the right one the relative difference between the Komar and ADM masses. The curves are labeled by the number of points in each dimensions.
}
\end{figure}

Using a spherical space, those equations have been solved using four different resolutions (9, 11, 13 and 17 points in each dimension). Sequences of black holes rotating at different values of $\Omega$ have been obtained. One single spherical shell is used and the value of $n_0$ is set to $0.5$ in all the cases (see Eq.(\ref{e:kerr_bclapse})). The results are presented as a function of the usual Kerr parameter $J/M^2= a/M$ where $J$ is the total angular momentum and $M$ the ADM mass of the system. Both quantities are computed by surface integrals at infinity (see Sec. \ref{ss:hkv}). The value of the Kerr parameter as a function of the adimensional quantity $M \Omega$ is plotted in Fig. \ref{f:a_f_ome}. A value of $a/M \approx 0.91$ is reached. This value is lower for lower resolutions as can be seen in Fig. \ref{f:kerr_errors}.

\begin{figure}[htb]
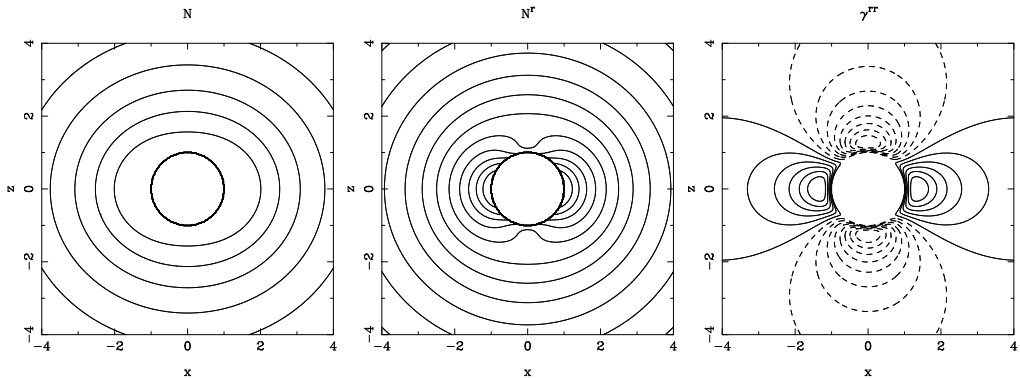

\includegraphics[width=0.32\textwidth]{lapse_kerr.eps} 
\includegraphics[width=0.32\textwidth]{shift_kerr.eps} 
\includegraphics[width=0.32\textwidth]{gamma_kerr.eps}
\caption{ \label{f:fields_kerr}
Values of the lapse function (left panel), the component $N^r$ of the shift (center panel) and of the component $\tilde{\gamma}^{rr}$ of the spatial metric. The associated value of $a/M$ is about $0.91$. The continuous (resp. dashed) lines show constant positive (resp. negative) value of the fields.
}
\end{figure}

Two error diagnostics are shown in Fig. \ref{f:kerr_errors}. On the left panel the residual error on the full set of equations is plotted. By full set, one means not only the equations used to solve the system but also the ones that were forgotten. In particular one can think of the $\left(\varphi, \varphi\right)$ component of Eq. (\ref{e:kerr_evol}) or the Dirac Gauge in whole space. This is a very strong test and it shows, as already stated, that the Dirac gauge is enforced in the whole space by only its $\theta$ component on the horizon. The right panel of Fig. \ref{f:kerr_errors} shows the relative difference between the ADM and Komar masses (as defined by Eqs. (\ref{e:komar}) and (\ref{e:adm})). Like in the binary case, those two quantities should be equal. However, it is never directly enforced in the single black hole case and so gives a check of the precision of the code. As stated in Sec.\ref{ss:hkv}, this is related to a virial theorem \cite{GourgGBM02}. Both plots in Fig. \ref{f:kerr_errors} are shown as a function of $a/M$ and for the four different resolutions. The errors decay as the resolution increases and are higher for higher values of the Kerr parameter. This is probably related to the fact that the fields have stronger gradients when $a/M$ increases and so need more points to be accurately described. This effect is also observed 
in \cite{VasseNJ09}. Some examples of the fields, for $a/M\approx 0.91$ are shown in Fig. \ref{f:fields_kerr}.

\section{Perspectives}\label{s:conclusion}

This presentation of the \K library was intended to convince the physicists that it can be a valuable tool in the study of a wide class of problems where partial differential equations are involved. Several domain decompositions have been presented (Sec. \ref{s:geometry}) and the associated spectral basis 
exhibited (Sec. \ref{s:spectral}). The discretization of field equations by means of the $\tau$ and Galerkin methods have been discussed in Sec. \ref{s:equations} and the solution of the resulting non-linear system explained in Sec. \ref{s:system}. The construction of a code that uses the \K library is briefly sketched in Sec. \ref{s:user}.

The resolution of four different problems has also been presented. Solution called vortons have been computed in Sec. \ref{ss:vortons}, using a standard \K setting with spherical domains. In Sec. \ref{ss:critic} the computation of a critic solution in the context of core collapse has been presented. This case illustrates the use of specialized domains and different variables in different regions of space. Section \ref{ss:hkv} explained the study of spacetimes containing  two black holes. It makes use of the bispherical coordinates implemented in \K. The ability of the library to deal with intricate sets of equations is clearly illustrated by Sec. \ref{ss:kerr}, where the Kerr black hole is recovered, in the 3+1 formalism of general relativity.

Even if \K is currently in a production state, where it can be used to study new and interesting physics, there are still some developments that come to mind. One can for instance think about implementing new geometries like true cylindrical coordinates (as opposed to the specialized ones used in Sec. \ref{ss:critic}) or deformed spheres. Such spheres have been extensively used in the {\tt LORENE} library to match the physical surfaces of deformed objects, like rotating neutron stars. This case is in fact a situation where the surface of the domains in not known beforehand but must be determined by the code. To do so, there will be need to have some additional unknowns that would describe the shape of the various domains. Those unknowns would be associated to additional equations constraining the shapes. In the case of neutron stars, one would for instance demand that the surface of the object is the surface on which the specific enthalpy vanishes. Given those techniques have already been successfully applied in many codes, they should work properly with \K. Geometries with more than three spatial dimensions could also be studied, in order to simulate things like brane worlds. Needless to say that in this last case, the resulting size of the system would be quite big and possibly difficult to handle.

It would be also interesting to test the behavior of \K on much bigger clusters than the one used so far. Several such clusters (with much more than $1000$ nodes) are available to the scientific community and there is hope to install \K on such machines. Another way to deal with very big matrices would be to use iterative techniques to invert the Jacobian. As stated in Sec. \ref{ss:parallel}, this solution is currently not retained in \K because of several difficulties (lack of generality and absence of simple preconditionning techniques). Nevertheless, a much more detailed study of those algorithms should be undertaken. A successful use of the iterative techniques would potentially result in a dramatic decrease of the resources needed to handle the Jacobian, thus making the library easier to use, especially in the three-dimensional case.

A major extension of \K would be the possibility to perform time evolutions. Even when spectral methods are used to discretize space, time is usually dealt with by means of  finite difference schemes. One of the reason for this lies in the fact that the interval of interesting time is usually not known in advance. Also the discretization with respect to time is usually believed to lead to smaller errors that the spatial ones, thus making the use of temporal spectral methods less appealing. If those mixed schemes (finite difference in time and spectral in space), could be almost directly used with \K, it is not clear what this will improve with respect to already existing codes. Indeed, for free evolutions (like in \cite{PazosTDKT09}) for which the only equations solved are hyperbolic ones, the whole \K machinery would not be used but to compute the sources. On the other hand, for constrained evolutions (like in \cite{BonazGGN04}), one would need to solve elliptic equations at each time-step, which would probably be impracticable for three-dimensional problems with \K. The library however, could be very useful in investigating schemes that are fully spectral, in time as in space. Results of this type are very few and are only available for very simple systems (see \cite{HenniA09} for an example with only one spatial dimension). For such fully spectral codes, time is just another coordinate that is treated in the same manner as the spatial dimensions. If this could in theory be solved by \K, there is a long road to compute complicated system evolutions in this framework (like the binary black holes for instance). Preliminary tests, using systems with symmetries (like rotating single objects or pure gravitational waves), would be in order to check whether such computations are doable or not.

Obviously all the work done would be quite pointless without applications to real physics problems and it is the main axis of future work. There are plans to apply the library to the computation of solutions in gauge field theories, like monopoles with or without gravity, vortons in more complicated theories than the one exposed in Sec. \ref{ss:vortons}. In this area of physics, the field of application is huge. \K has been designed especially with this kind of things in mind and it is believed that it will be very efficient in findings those solutions. 

Another area of application is the one of general relativity, especially in the context of compact objects, black holes or neutrons stars. One of the main application would be the computation of binary black holes configurations without the conformal flatness approximation. In a sense it would be a merging of Sec. \ref{ss:hkv} and \ref{ss:kerr}. But there are some conceptual difficulties coming from the presence of outgoing gravitational waves in this case. The applications of \K to cases containing neutron stars would also be interesting. If the gravitational field is usually weaker than for black holes and so easier to handle, the inclusion of matter can cause many additional difficulties.

In the hopefully long life of \K, there is hope that it will also be applied to problems that are not yet known to the author. This would fully test the modularity of \K. Should this be successful, then \K would have reached its main objective.

\end{document}